\documentclass{article}

\usepackage{arxiv}

\usepackage[utf8]{inputenc} 
\usepackage[T1]{fontenc}    
\usepackage{hyperref}       
\usepackage{url}            
\usepackage{booktabs}       
\usepackage{amsfonts}       
\usepackage{nicefrac}       
\usepackage{microtype}      
\usepackage{lipsum}		
\usepackage{graphicx}
\usepackage{natbib}
\usepackage{doi}
\usepackage{amsmath}

\title{Static and Dynamic Functional Brain Networks Reveal Distinct Connectivity Alterations Across the Alzheimer's Disease Spectrum}

\author{{\hspace{1mm}Nicolas Rubido$^\&$}\\
	Aberdeen Biomedical Imaging Centre\\
	University of Aberdeen\\
    \And
    {\hspace{1mm}Venia Batziou$^\&$}\\
    Health Data Science\\
	Swansea University Medical School\\
	Swansea University \\
    \And
    {\hspace{1mm}Marwan Fuad}\\
    Health Data Science\\
	Swansea University Medical School\\
	Swansea University \\
	\And
	{\includegraphics[scale=0.06]{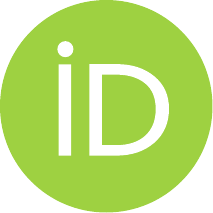}\hspace{1mm}Vesna Vuksanovi\'c} \thanks{correspondence: vesna.vuksanovic@swansea.ac.uk \\ $^\&$ These authors contributed equally} \\
	Health Data Science\\
	Swansea University Medical School\\
	Swansea University \\
Correspondence:\texttt{vesna.vuksanovic@swansea.ac.uk}\\
}

\renewcommand{\shorttitle}{Dynamic Functional Connectivity in AD and MCI}

\hypersetup{
pdftitle={A template for the arxiv style},
pdfsubject={q-bio.NC, q-bio.QM},
pdfauthor={David S.~Hippocampus, Elias D.~Striatum},
pdfkeywords={First keyword, Second keyword, More},
}

\begin{document}


\maketitle
\begin{abstract}
Brain network analysis using functional MRI has advanced our understanding of cortical activity and its disruption in neurodegenerative disorders underlying dementia. Recently, research has focused on dynamic (time-varying) brain networks that capture both spatial and temporal patterns of regional cortical co-activity. However, this approach remains relatively unexplored across the Alzheimer’s disease spectrum. In this study, we analysed age- and sex-matched static and dynamic functional brain networks derived from resting-state fMRI data in 315 individuals with AD, mild cognitive impairment (MCI), and cognitively normal healthy controls (HC) from the ADNI-3 cohort. Functional networks were constructed using the Juelich brain atlas, with static connectivity estimated from full time series and dynamic connectivity derived using a sliding-window approach. Group differences were assessed at both the link and node levels using non-parametric statistics and bootstrap resampling. Finally, time series extracted from regions identified by the primary connectivity analyses were used in an exploratory Random Forest classification framework to assess whether these network alterations possessed discriminative information at the individual level. HC and MCI showed largely similar node-level static and dynamic connectivity patterns, whereas AD exhibited pronounced alterations in both network organisation and temporal variability. We identified stable (stationary) differences in functional connectivity between white matter regions and parietal and somatosensory cortices, whereas temporally varying differences were consistently observed in connections involving the amygdala and hippocampal formation. In addition, node centrality analysis suggested that white matter connectivity differences are predominantly local in nature. Our results highlight shared and unique functional connectivity patterns in both static and dynamic functional networks, emphasising the importance of incorporating dynamic information into brain network analyses of the Alzheimer's spectrum. Furthermore, Random Forest classification demonstrated that network-derived features identified from static and dynamic connectivity analyses contain information relevant to distinguishing clinical groups at the individual level.
\end{abstract}

\maketitle

\section*{Introduction}
\label{sec_Intro}

Brain networks derived from functional magnetic resonance imaging (fMRI) have contributed substantially to our understanding of functional connectivity, defined as coordinated patterns of activity between spatially distributed brain regions \citep{sporns2010networks, fornito2016fundamentals, bijsterbosch2017introduction}. Mapping functional networks from neuroimaging data has provided important insights into the neural mechanisms underlying neurological and psychiatric disorders \citep{mattson2000apoptosis, zecca2004iron, pievani2011functional, stam2014modern}. In Alzheimer's disease (AD), functional connectivity differs from that observed in cognitively healthy individuals \citep{eguiluz2005scale, damoiseaux2006consistent, honey2009predicting, bullmore2009complex, bullmore2012economy}. These alterations have been linked to the accumulation of amyloid-$\beta$ and tau pathology, as well as to structural white matter damage \citep{borne2024interplay}. Network abnormalities have also been associated with the severity of cognitive and behavioural symptoms \citep{stam2007small, binnewijzend2014brain, luo2017intrinsic, dai2019disrupted, lei2019longitudinal}. Network-based studies further suggest that AD does not affect the brain uniformly, but preferentially disrupts highly connected hub regions within temporal, parietal, and frontal association cortices that support higher-order cognitive functions \citep{he2008structural, stam2009graph}. \par 
AD is increasingly recognised as a disorder of large-scale brain network disruption, often described as a disconnection syndrome \citep{delbeuck2003alzheimer}. Similar, although generally less pronounced, network alterations have been reported in individuals with mild cognitive impairment (MCI), a condition associated with an increased risk of progression to dementia \citep{zhang2011multimodal, gilligan2019no}. Previous studies have demonstrated abnormalities in both local and global network organisation, including altered clustering, reduced network efficiency, disrupted hub architecture, changes in modular organisation, and impaired connectivity within the default mode network \citep{de2012disruption, heringa2014multiple, lo2010diffusion, yao2010abnormal, reijmer2013disruption, liu2012altered, wang2020resting, chen2013modular, contreras2019resting, vuksanovic2019cortical, greicius2004default, petrella2011default, agosta2012resting, dillen2017functional}. Collectively, these findings indicate that AD affects multiple aspects of functional brain organisation and that such alterations can be detected before the onset of dementia. \par 
Most resting-state fMRI studies quantify connectivity using correlations calculated across the entire scan, producing a single static representation of functional interactions. This approach assumes that connectivity remains relatively stable throughout the acquisition period. However, growing evidence suggests that large-scale brain networks exhibit substantial temporal variability even during rest \citep{Vuksanovic2014, alderson2020metastable}. Dynamic functional networks (dFNs) aim to capture these fluctuations by characterising connectivity within shorter temporal windows, thereby revealing transient patterns of network organisation that are averaged out in conventional static analyses. Analysis of dynamic functional connectivity has shown promise for identifying clinically relevant changes in brain disorders \citep{filippi2019resting,moguilner2021dynamic}. For example, reduced network flexibility has been reported in temporal lobe epilepsy and linked to impaired language function \citep{he2018disrupted}, while studies in schizophrenia have demonstrated altered occupancy of dynamic connectivity states involving default mode and language networks \citep{lee2018resting}. These findings suggest that dynamic approaches may provide complementary information about brain dysfunction beyond that obtained from static connectivity alone.
In Alzheimer's disease, dynamic functional connectivity has attracted increasing interest \citep{rubido2024genetic, cordova2017disrupted, quevenco2017memory, nunez2021abnormal, sendi2021link}. Several studies have reported that dynamic measures improve the differentiation of AD from healthy ageing and may be more sensitive to cognitive impairment than conventional static connectivity estimates \citep{quevenco2017memory, du2018classification}. Nevertheless, relatively few studies have directly compared static and dynamic functional network organisation across the Alzheimer's disease spectrum, and the extent to which dynamic analyses provide information beyond that captured by static connectivity remains incompletely understood. \par 
In this study, we investigated large-scale functional brain networks in individuals with AD, MCI, and healthy ageing using resting-state fMRI data from the ADNI cohort. We performed a comparative analysis of static and dynamic functional networks to determine whether temporal fluctuations in connectivity provide additional information regarding disease-related network disruption. We further evaluated whether network features identified through static and dynamic analyses could distinguish AD and MCI from healthy controls. We hypothesised that dynamic functional networks would reveal alterations not captured by conventional static connectivity measures and would provide improved sensitivity to disease-related network abnormalities across the Alzheimer's disease spectrum.

\section*{Methods}
\subsection*{Participants and Cohorts} Data used in this study were obtained from the Alzheimer's Disease Neuroimaging Initiative  \href{https://adni.loni.usc.edu/}{(ADNI)}  data base. We used demographic, clinical and MRI data from $N = 315$ individuals participating in the study. The main inclusion criterion was that the fMRI datasets were acquired using the same resting-state protocols (see Sec.~\hyperref[sec:dataanalysis]{Data Analysis} for details). The $315$ participants were selected from three groups: Healthy Controls (HC), Mild cognitive Impairment (MCI), and Alzheimer's Disease (AD) groups. Demographic and clinical characteristics for each group are summarised in Table~\ref{table:studygroups}. 

\subsection*{Imaging Datasets and Processing} We analysed resting-state functional (rs-fMRI) images from 315 participants. fMRI were acquired using ADNI-3's basic EPI-BOLD protocol (details at \href{http://adni.loni.usc.edu/data-samples/data-types/mri}{ADNI's website}) and in \citep{weiner2017alzheimer}. In short, the scanning time of each participants was for up to $10$ minutes using the same two-time accelerated $3~T$ scanner, following an even/odd interleaved axial-slicing (inferior to superior) of $3.4\,mm$ with $(3.4375\,mm)^2$ pixels (FOV = $220\times220$; $P>>A$ phase encoding; TE = $30\,ms$; TR = $3~s$). 

Image preprocessing, which includes brain extraction, registration in standard MNI space, and brain tissue segmentation was carried out using FMRIB's pipeline \textit{fsl\_anat} in its default settings. Pre-processing of fMRI is done by applying FMRIB's Expert Analysis Tool, \textit{FEAT}, resulting in Blood-Oxygen-Level-Dependent (BOLD) signals of $N_T = 197$ data points ($197\times$~TR acquisition times) for each voxel. Please see the Supplementary Information for a more detailed description of the image processing and the steps applied, including quality control for image processing.   
\begin{table}[h!]
 \caption{Participants' demographic and clinical characteristics.}
\begin{center}
\begin{tabular}{@{}cccc@{}}
     \hline
     {\bf } & {\bf HC} & {\bf MCI} & {\bf AD} \\ [0.5ex] 
     \hline
     Participants & $141$ & $128$ & $46$ \\ 
     Age (s.d.) & $79.5$ ($5.6$) & $76.8$ ($4.7$)$^\star$ & $78.5$ ($5.9$) \\ 
     Sex (F:M) & $68:73$ & $52:76$ & $17:29$ \\ [0.5ex] 
     \hline\hline
     {\bf Clinical Scores} &   &   &  \\ [0.5ex] 
      \hline
     MMSE  & $28.2$ ($3.6$) & $27.6$ ($2.5$) & $22.4$ ($3.9$)$^{\star,\,\star\star}$ \\
     \hline
    \end{tabular}
    \label{table:studygroups}
    \end{center}
\tiny{Abbreviations: M-male, F-Female, s.d.-standard deviation, HC-healthy controls, AD-Alzheimer's Disease, MCI-Mild Cognitive Impairment, $^\star$ significant difference ($p < 0.01$) from HC. $^{\star\star}$ significant difference ($p < 0.01$) from MCI.}
\end{table}

\subsection*{Functional Connectivity Analysis}

Cortical regions of interest were defined using the Jülich Brain Atlas (JBA) \citep{amunts2020julich}, which parcellates the cortex into 121 regions (Supplementary Information). To evaluate the robustness of the findings to anatomical parcellation, all principal analyses were repeated using the Harvard-Oxford Brain Atlas (H-OBA) (\href{https://neuinfo.org/data/record/nlx_144509-1/RRID:SCR_001476/resolver/pdf&i=rrid:scr_001476}{Harvard - Oxford Brain Atlas [RRID:SCR$\_$001476]}), which comprises 48 cortical regions. The JBA was used for all primary analyses, while results obtained using the H-OBA are presented in the Supplementary Information. Consistent findings across both atlases indicated that the observed effects were not dependent on the choice of brain parcellation.

For each participant, a representative BOLD signal was obtained for every region by averaging the resting-state fMRI time series across all voxels within that region. Functional connectivity was estimated using Pearson correlations between regional time series, resulting in weighted undirected connectivity matrices for each participant. 

\subsection*{Static Functional Connectivity}

Static functional connectivity was calculated using the entire resting-state time series. For each participant, connectivity matrices were generated by computing correlations between all pairs of cortical regions. Significant correlations were retained as weighted network connections, with connection strength corresponding to the magnitude of the correlation.  Only statistically significant correlations FDR-corrected (q$<$0.05) were retained to reduce the influence of spurious associations. Using this threshold, approximately 39\% of weaker connections were excluded from the JBA-derived static networks.

\subsection*{Dynamic Functional Connectivity}

Dynamic functional connectivity was assessed using a sliding-window approach. This method allows functional interactions between brain regions to vary over time rather than assuming that connectivity remains constant throughout the scan. To capture network dynamics over different temporal scales, analyses were performed using window lengths of 60, 120, and 180 seconds. These window lengths were selected to sample short-, intermediate-, and longer-timescale fluctuations while maintaining sufficient data within each window for reliable connectivity estimation  \citep{leonardi2015spurious, zalesky2015towards}.  Windows overlapped by 50\%, generating a sequence of connectivity matrices across the scan. As a result, each pair of brain regions was characterised by a temporal profile of connectivity rather than a single static value. For example, the shortest window length generated 18 partially overlapping windows across the scanning session.

To reduce the influence of unstable short-window correlations, dynamic connectivity was calculated relative to each participant's corresponding static functional connectivity network. Specifically, a dynamic link between two regions was retained only when it was statistically significant (p$<$0.01) different from the corresponding static link. This procedure was used to identify dynamic changes in functional connectivity that differ from an individual's baseline (static) connectivity. The resulting dynamic networks therefore represent temporally localised connectivity events that differ from each participant's stationary connectivity profile. On average, approximately 30\% dynamic connections (links) were excluded using this procedure for the JBA and 28\% for the H-OBA. The remaining connections were used to characterise dynamic network organisation and its variation across clinical groups.

\subsection*{Functional Networks' Metrics}
We characterised functional connectivity in three study
groups based on the importance of the network nodes.
A straightforward method to assess the importance of a
network node is to compute its centrality. The centrality of
a node is a measure that quantifies how important or influential a node is within a network. Centrality can be expressed in
various ways; thus, there are multiple types of centrality
measures. We characterised the importance of each node in
the dFNs and sFN of each individual in the cohort by calculating the node strength and eigenvector centrality. Both centrality measures, node strength and eigenvector centrality, were calculated on the same number of links (network density) across all individual networks. This was achieved by applying a threshold that ensures equal network density across participants.

\paragraph{\textbf{Node Strength.}} To characterise the importance of individual brain regions within the functional network, we calculated \textit{node strength}, a weighted measure of network centrality. Unlike degree centrality, which counts the number of connections associated with a node, node strength takes into account the magnitude of those connections and therefore reflects the overall level of functional coupling between a brain region and the rest of the network. For each participant, node strength was calculated from the weighted functional connectivity matrices derived from the resting-state fMRI data. Higher node strength values indicate that a region exhibits stronger overall functional interactions with other regions and may therefore play a more prominent role in network communication. For dynamic functional networks, node strength was computed separately within each sliding window, providing a time-resolved measure of regional network participation. Specifically, node strength was defined as the sum of all functional connections associated with a given node: \begin{equation} \kappa_{t_m}^{(s)}(i)= \sum_{j=1,\;j\neq i}^{N_{ROI}} r_{t\in[t_m,t_m+\Delta t]}^{(s)}(i,j), \label{eq_NodeStrength} \end{equation} where $r_{t\in[t_m,t_m+\Delta t]}^{(s)}(i,j)$ denotes the functional connectivity between regions $i$ and $j$ for participant $s$ within the sliding window $[t_m,t_m+\Delta t]$, and $N_{ROI}$ is the total number of brain regions in the atlas. Thus, node strength quantifies the cumulative connectivity of each region with the remainder of the network during a given time window.

\paragraph{\textbf{Eigenvector Centrality.}}To complement node strength, we calculated \textit{eigenvector centrality}, a measure that captures not only how strongly a brain region is connected to other regions, but also how influential those neighbouring regions are within the network. Consequently, regions connected to other highly connected or highly influential regions receive higher eigenvector centrality values than regions connected primarily to less influential nodes. In the context of functional brain networks, eigenvector centrality identifies regions that are positioned within the core of the network and may therefore play an important role in supporting large-scale communication. Unlike node strength, which reflects the overall magnitude of a region's direct connections, eigenvector centrality additionally considers the broader network context in which those connections occur. For each participant, eigenvector centrality was calculated from the weighted functional connectivity matrices obtained from resting-state fMRI data. For dynamic analyses, eigenvector centrality was estimated separately within each sliding window, providing a time-resolved measure of the influence of each brain region on network organisation. Formally, eigenvector centrality is defined as the principal eigenvector of the connectivity matrix: \begin{equation} e_{t_\Delta}(i) = \frac{1}{\lambda_{t_\Delta}} \sum_{j=1}^{N_{ROI}} r_{t_\Delta}(i,j)\, e_{t_\Delta}(j), \end{equation} where $r_{t_\Delta}(i,j)$ denotes the functional connectivity between regions $i$ and $j$ during the time window $t_\Delta$, $\lambda_{t_\Delta}$ is the largest eigenvalue of the corresponding connectivity matrix, and $e_{t_\Delta}(i)$ represents the eigenvector centrality of region $i$. Higher values indicate regions that are connected to other highly influential regions and therefore occupy more central positions within the network. Thus, while node strength reflects the overall level of connectivity of a brain region, eigenvector centrality provides a more global measure of network influence by considering both direct and indirect patterns of connectivity throughout the network.

\subsection*{Classification Analysis} Random Forest (RF) classification was used to evaluate whether connectivity alterations identified at the group level contained sufficient information to discriminate between individuals with AD, MCI and HC groups. The purpose of these analyses was to determine whether network features associated with disease-related connectivity changes at the group level, also show discriminative value at the individual level. \par 
Classification models were trained using either the full resting-state fMRI time series (197 time points) or shorter temporal segments corresponding to the dynamic connectivity analyses (60, 120, and 180 seconds). The brain regions were selected based on the results of the static and dynamic functional connectivity analyses. The classification analyses were hypothesis-driven and intended
to evaluate the discriminative value of network alterations identified in the primary connectivity analyses. Three sets of regions were investigated: (i) hippocampal and amygdala regions (nodes 7--26), which were primarily associated with dynamic connectivity alterations; (ii) white matter regions (nodes 93--115), which showed prominent static connectivity differences; and (iii) a subset of white matter regions (nodes 96--113) that consistently differentiated groups across both local and global network measures (Figs.~\ref{fig:nodestrength_groups}~and~\ref{fig:eig_groups}). For each selected region, the resting-state fMRI time series was used as input to the classifier, with individual time points treated as features. Clinical group labels (AD, MCI, or HC) served as the target variable. Classification performance was evaluated using an RF model comprising 100 decision trees and stratified 10-fold cross-validation, ensuring that class proportions were maintained across folds.

\subsection*{Statistical Analysis}
The primary aim of the statistical analyses was to determine whether dynamic measures of functional connectivity provide complementary information to conventional static connectivity measures across the Alzheimer's disease spectrum. Network differences were evaluated at both the regional (node) and connection (link) levels. Network measures were used to characterise network organisation at the group level, whereas classification analyses were performed to evaluate whether group-level network alterations possessed discriminative value at the individual level. \par \textbf{Node-Level Analysis} Group differences in demographic characteristics and network metrics, including node strength and eigenvector centrality, were assessed using the non-parametric Kruskal--Wallis test. When significant group effects were observed, pairwise comparisons between clinical groups were performed using the Wilcoxon rank-sum test. Statistical significance was set at $p < 0.01$. \par
\textbf{Link-Level Analysis} Differences in functional connectivity at the link level were evaluated using the Kruskal--Wallis test applied to individual connection strengths across the three study groups. Given the large number of possible connections, a bootstrap-based procedure was used to assess the robustness of the findings. Specifically, 100 independent bootstrap realisations were generated. In each realisation, 150 participants were randomly selected, comprising 50 individuals from each clinical group (AD, MCI, and HC). Sampling preserved the sex distribution of the AD group to maintain demographic consistency across bootstrap samples. For each connection, statistical comparisons were performed across the selected participants using the Kruskal--Wallis test. When significant overall group differences were identified, post hoc pairwise comparisons were conducted using the Wilcoxon rank-sum test. To account for multiple comparisons, false discovery rate (FDR) correction was applied across all tested connections, with an adjusted significance threshold of $q < 0.05$. \par \textbf{Classification Analysis} The ability of network-derived features to distinguish between AD, MCI, and HC participants was evaluated using Random Forest classification. Classification analyses were considered exploratory and were performed to assess whether network alterations identified at the group level also contained information useful for individual-level discrimination.

\section*{Results}\label{sec_Results}
\subsection*{The Node-Level Analysis: Node Strength and Eigenvector Centrality} 
\begin{figure}
    \centering
    \includegraphics[width=0.85\columnwidth]{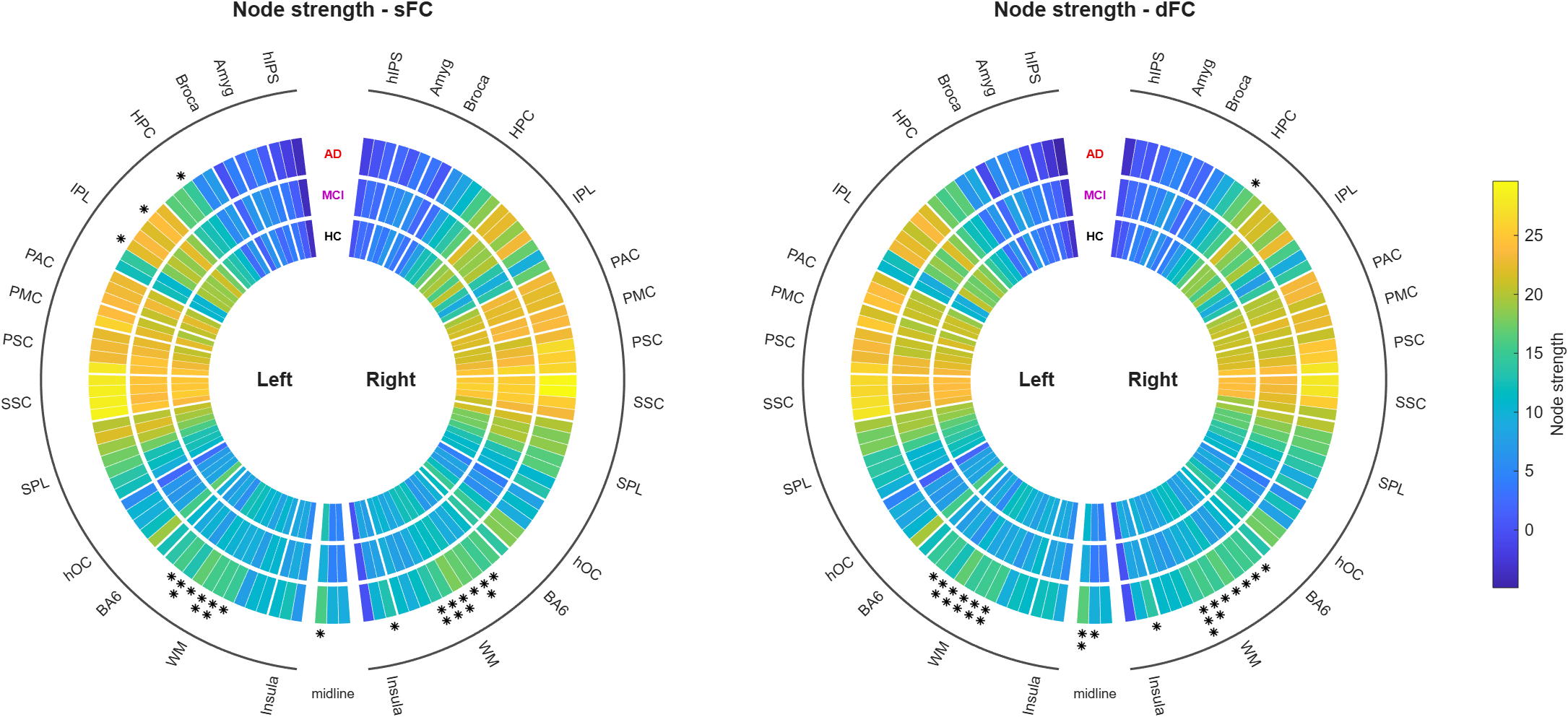}
\caption{\textbf{Node strength in static and dynamic functional networks.} Ring plots show node strength for the 121 regions of the Juelich Brain Atlas in healthy controls (HC), mild cognitive impairment (MCI) and Alzheimer's Disease (AD) groups. Each wedge is one region; regions are grouped into anatomical systems, marked by the outer labels. The three concentric rings show the three cohorts. Colour indicates node strength on a common scale for all rings. Asterisks mark regions with significant differences between HC and AD ( *: 0.001$<p<$0.01; **: $10^{-4}<p<$0.001; **: $p<10^{-4}$ ).}

    \label{fig:nodestrength_groups}
\end{figure}

Figure~\ref{fig:nodestrength_groups} presents node strength and eigenvector centrality for sFNs and dFNs across the HC, MCI, and AD groups. Static connectivity summarises average interactions across the scanning session, whereas dynamic connectivity characterises temporal fluctuations in these interactions. Both approaches identified disease-related differences; however, dynamic connectivity revealed additional alterations involving limbic and association regions that were not detected consistently in the static networks. \par 
Overall, node strength exhibited similar patterns across sFNs and dFNs, with only modest differences between the two approaches. The most consistent between-group differences were observed in white matter regions, where several effects were present in both static and dynamic networks (asterisks in Figure~\ref{fig:nodestrength_groups}). A subset of regions showed significant differences only in the dynamic networks. These included nodes within the hippocampal formation, inferior parietal lobule, and insular cortex, indicating that connectivity alterations in these regions were more apparent when temporal variability in network interactions was considered. \par
\begin{figure}[h!]
    \centering

    \includegraphics[width=0.85\columnwidth]{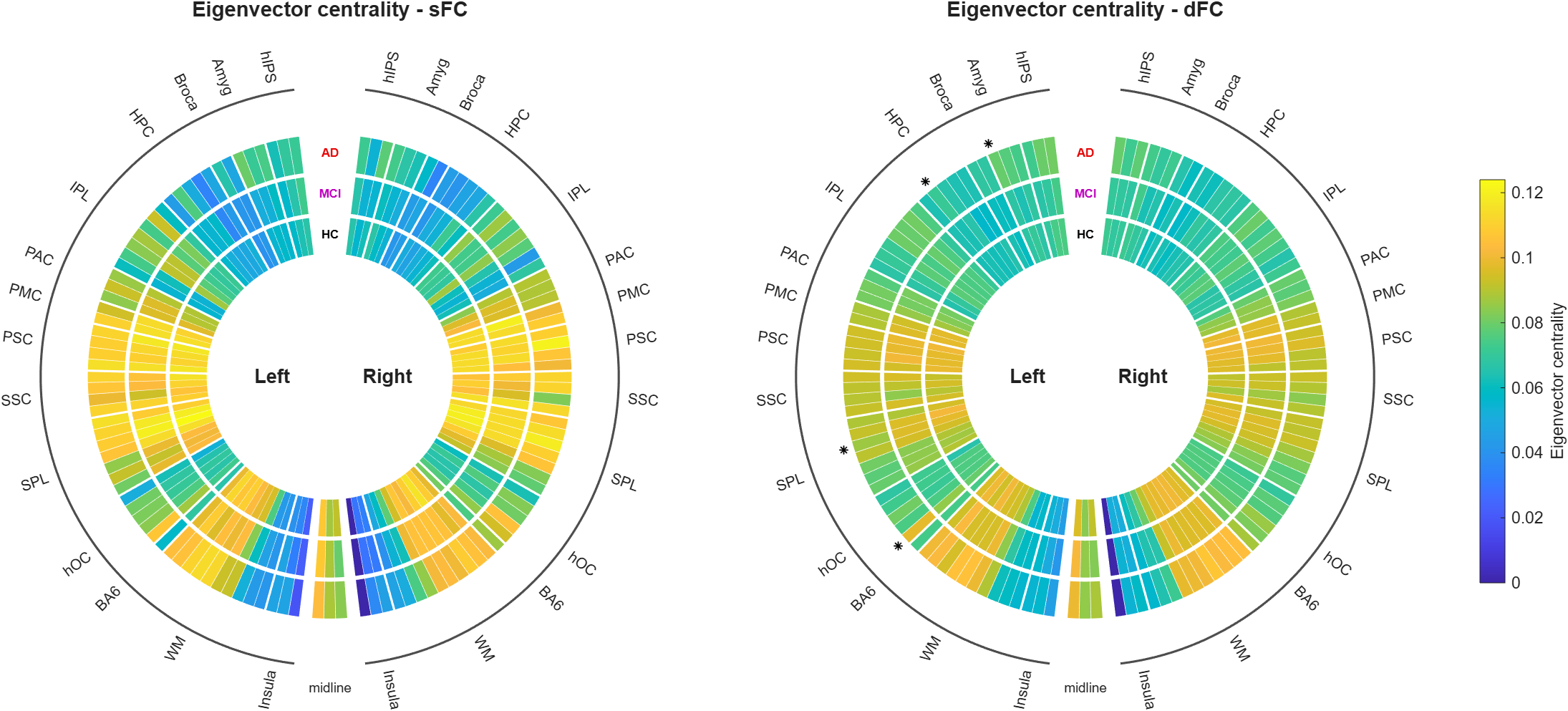}
    \caption{\textbf{Eigenvector centrality in static and dynamic functional networks.} Ring plots show eigenvector centrality for the 121 regions of the Juelich Brain Atlas in healthy controls (HC), mild cognitive impairment (MCI) and Alzheimer's Disease (AD) groups. Each wedge is one region; regions are grouped into anatomical systems, marked by the outer labels. The three concentric rings show the three cohorts. Colour indicates eigenvector centrality values on a common scale for all rings. Asterisks mark regions with significant differences between HC and AD ( *: $<p<$0.001).}

    \label{fig:eig_groups}
\end{figure}

 
Eigenvector centrality (Fig.~\ref{fig:eig_groups}) showed a broadly similar distribution across sFNs and dFNs. Statistically significant group differences, however, were observed primarily in the dynamic networks and involved only four nodes; the amygdala, hippocampus, superior parietal lobule, and higher-order visual cortex. In contrast to the node strength results, white matter regions did not show significant differences in eigenvector centrality. This pattern indicates that the observed white matter effects were more prominent at the level of direct regional connectivity than in measures reflecting the global influence of a node within the network.

\begin{table*}
\caption{The accuracy of the Random Forest classifier on the entire time series of 197 time points and on windows of $\Delta$t = 60, 120 and 180 seconds, reported as the mean $\pm$ standard error. The classifier was trained using 10-fold cross-validation based on 100 decisions trees. For the nodes names see the SI Table~\ref{tab:JA_regions}.}
\begin{center}
{\begin{tabular}{@{}cccccc@{}}
\hline
\toprule
\multicolumn{5}{r}{{\bf Accuracy ($\%$)}} \\ 
     {\bf Nodes (Idx)} & {\bf Groups} & T= 197 & $\Delta$t = 60 & $\Delta$t= 120 & $\Delta$t = 180 \\ [0.5ex] 
       \hline
     7 -- 26 & MCI vs AD & 50.6$\pm$3.1 & 52.8$\pm$1.7 & 53.0$\pm$4.1 & 58.8$\pm$3.1 \\ 
     ~ & MCI vs HC & 73.6$\pm$.8 & 73.6$\pm$.8 & 73.1$\pm$1.2 & 73.6$\pm$1.1\\
     ~ & HC vs AD & 75.5$\pm$.7  & 75.4$\pm$.7 & 75.4$\pm$.7 & 79.4$\pm$1.4 \\
     ~ & HC + MCI + AD  & 46.4$\pm$2.2  & 45.4$\pm$3.2 & 44.8$\pm$3.0 & 57.8$\pm$2.5 \\
     \hline
           93 -- 115 & MCI vs AD & 73.6$\pm$.8 & 73.6$\pm$.8 & 73.6$\pm$.8 & 74.2$\pm$1.1 \\ 
       ~& MCI vs HC & 52.6$\pm$.3 & 53.1$\pm$1.7 & 53.2$\pm$2.1 & 71.8$\pm$2.0 \\ 
       ~& HC vs AD & 75.4$\pm$.7 & 75.4$\pm$.7 & 75.4$\pm$.7 & 78.7$\pm$1.2 \\ 
       ~& HC + MCI + AD & 45.8$\pm$2.6 & 47.6$\pm$2.1 & 50.0$\pm$1.5 & 63.0$\pm$2.5 \\ 
     \hline
     
     96, 97, 101:109, 112, 113 & HC vs AD & 75.5$\pm$0.8 & 75.5$\pm$0.8 & 75.5$\pm$0.8 & 76.5$\pm$1.1 \\ 
     ~ & HC vs MCI & 55.5$\pm$2.1 & 49.1$\pm$2.1 & 53.2$\pm$2.4  & 68.5$\pm$2 \\ 
     ~ & AD vs MCI & 45.4$\pm$3 & 50.2$\pm$3.1 & 42.5$\pm$2.5  & 67.5$\pm$2 \\ 
      ~& HC vs (AD + MCI) & 54.0$\pm$2.0 & 47.6$\pm$2.0 & 52.7$\pm$2.6 & 74.3$\pm$3.2\\ 
     \hline
    \end{tabular}}
\label{table:rf_197}
\end{center}
\end{table*}

\subsection*{Random Forest Classification} Table~\ref{table:rf_197} summarises the performance of the Random Forest classifier trained using regional BOLD time series data. Three sets of regions were used as input features: (i) nodes 7--26 (amygdala and hippocampal formation), identified primarily through dynamic network analyses; (ii) nodes 93--115 (white matter regions), which showed prominent differences in static connectivity; and (iii) a subset of white matter nodes (96--113), identified consistently across both local and global network measures (Figs.~\ref{fig:nodestrength_groups}~and~\ref{fig:eig_groups}). \par 
Binary classification generally outperformed three-class classification. The highest accuracy ($\approx79\%$) was achieved for HC versus AD using the amygdala--hippocampal node set and a window length of $\Delta t = 180$\,s. White matter node sets also yielded good discrimination between HC and AD. In contrast, classification performance for MCI versus AD was consistently lower across all feature sets and time windows. Classification performance improved with increasing temporal window length. Across all node sets, the highest accuracies were typically obtained using the longest window ($\Delta t = 180$\,s). \par 
Three-class classification (HC, MCI and AD) showed lower performance than pairwise comparisons, reaching a maximum accuracy of 63.0\% using white matter regions and a 180-second window. Classification performance generally improved with increasing window length, with the highest accuracies obtained using 180-second windows. These results suggest that regional BOLD signals selected based on network alterations contain information relevant to distinguishing clinical groups, although the classification findings should be considered exploratory. \par

\subsection*{Link-Level Analysis of Static Functional Connectivity} 
\begin{figure}
  \centering
   \includegraphics[width=0.48\columnwidth]{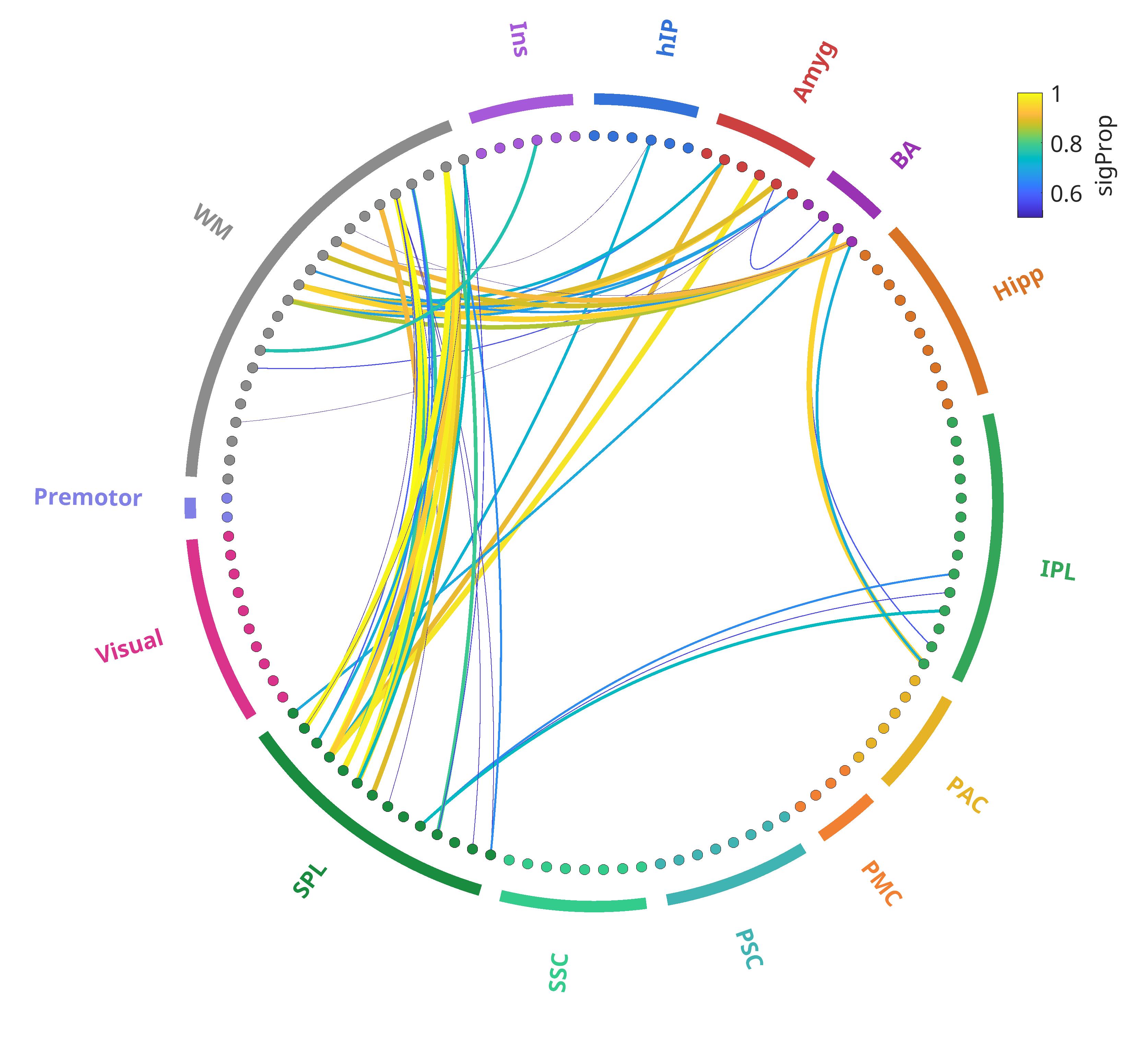}
    \caption{\textbf{Static functional network differences at the link level across the three study groups.}
Links (i.e., pairwise correlations) between nodes of the J{\"u}lich Brain Atlas are shown when they pass the Kruskal–Wallis (K-W) tests for group differences (HC, MCI, AD). For each node pair $(i,j)$, statistical significance was evaluated using 100 independent resampling realisations, each consisting of 150 participants (50 per group). Colour coding represents the portion of each link passed the K-W test ($p<0.01$). This procedure ensures that reported differences reflect robust group effects across balanced cohort samples. A full list of regions, their labels, indices and abbreviations can be found in Supplementary Table~\ref{tab:JA_regions}.}
     \label{fig:static_link_ring_groups}
\end{figure}

Figure~\ref{fig:static_link_ring_groups} summarises group differences in individual functional connections within the static networks. Significant differences were identified predominantly in connections linking white matter regions with the superior parietal lobule, amygdala, and hippocampal formation. These effects were consistently observed across bootstrap realisations, indicating stable group-level differences in static functional connectivity. The strongest and most reproducible alterations involved limbic and parietal connectivity pathways (Figure~\ref{fig:static_link_ring_groups}; Supplementary Figure~\ref{figsi:ring_KW_sigProp_thresholds}).

\begin{figure}
    \centering
     \includegraphics[width=0.85\columnwidth]{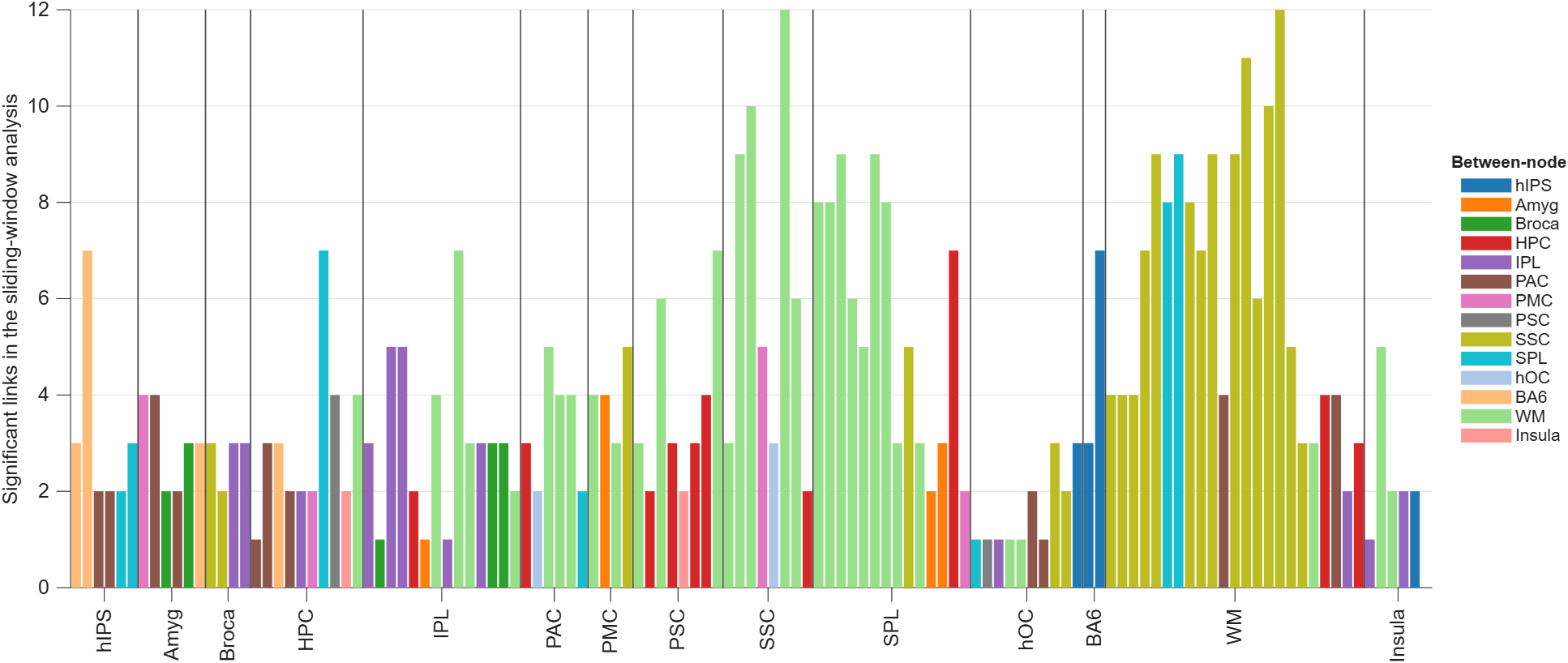}
    \caption{ {\bf Temporal persistence of group differences in functional connectivity}. The bar height represents the number of sliding windows (out of 18) in which the region's strongest link was identified as statistically significant based on the realisation-averaged, FDR-corrected p-values. Bar colour indicates the anatomical system of the region at the other end of that link. For each sliding window a group-level Kruskal-Wallis test was performed on each link across the three cohorts. A link was considered significant in a window if its realisation-averaged FDR-corrected p-value passed the significance threshold (q $<$ 0.05). The most persistent group differences involve links between somatosensory (SSC) or superior parietal (SPL) regions and white matter (WM) structures, which remain significant in up to 12 windows of the 18 in total. A full list of regions, including their labels, indices, and abbreviations, is provided in Supplementary Table~\ref{tab:JA_regions}.}
 \label{fig:sig_windows}
\end{figure}

Although significant overall group effects were identified at the link level, post hoc pairwise comparisons did not survive correction for multiple testing. This finding is consistent with small but widespread connectivity alterations distributed across the Alzheimer's disease spectrum, rather than a limited number of highly discriminative connections. To explore the direction of these effects, we subsequently examined deviations in static and dynamic connectivity patterns in AD and MCI relative to healthy controls. 

\subsection*{The Link-Level Analysis: Dynamic Functional Networks} \label{sec:dFNinter-cohort-median} Figure~\ref{fig:sig_windows} summarises the temporal persistence of group differences in dynamic functional connectivity across the AD, MCI, and HC cohorts. For each brain region, the height of the bar indicates the number of sliding windows in which its strongest functional connection showed a significant group effect following FDR correction. This analysis therefore identifies regions whose connectivity patterns consistently distinguish the three cohorts over time. The most persistent effects were observed in WM regions and their interactions with somatosensory and parietal cortices. Several WM regions remained significant in up to 12 of the 18 sliding windows, indicating that connectivity differences involving these regions were present throughout a large proportion of the scan. The strongest and most persistent effects involved links between WM structures and the somatosensory cortex or superior parietal lobule, suggesting that alterations within these systems represent a stable feature of functional network disruption across the Alzheimer's disease spectrum. \par 
Beyond WM-related effects, recurrent group differences were also observed in the SPL, SSC, hippocampal formation, amygdala, and inferior parietal lobule (IPL), although these effects were generally present in fewer sliding windows. In contrast, regions such as the higher-order visual cortex (hOC) showed only sporadic differences, indicating less stable temporal effects. Overall, the dynamic link-level analysis revealed that group differences were not distributed uniformly across the brain. Instead, they were concentrated within a subset of white matter, sensorimotor, parietal, and limbic regions. The persistence of these effects across multiple sliding windows suggests that they reflect reproducible alterations in network organisation rather than isolated transient fluctuations. \par 
To further characterise the direction of these connectivity changes, group-level median static and dynamic connectivity patterns were compared with healthy controls (Supplementary Information). While static analyses highlighted robust differences involving white matter and sensorimotor networks, dynamic analyses additionally revealed temporally varying alterations involving limbic and association regions, supporting the complementary information provided by time-varying connectivity measures.

\section*{Discussion} The present study compared static and dynamic functional connectivity approaches for characterising large-scale network alterations across the Alzheimer's disease spectrum. Using resting-state fMRI data from the ADNI cohort, we identified significant disruptions in network organisation in AD, with more limited abnormalities in MCI. While both static and dynamic analyses detected disease-related changes, dynamic connectivity revealed additional alterations involving limbic and association networks that were not consistently detected by static connectivity measures. These findings suggest that temporal variations in functional interactions provide complementary information regarding network dysfunction associated with neurodegeneration. \par  One of the main findings of this study was the prominent involvement of hippocampal- and amygdala-related connectivity. These structures play key roles in memory formation, emotional processing, and the integration of cognitive information, all of which are affected in Alzheimer's disease. Most previous resting-state studies have focused on changes in average connectivity strength within memory-related networks. In contrast, our results indicate that disease-related abnormalities are not restricted to mean connectivity but also involve changes in the temporal organisation of communication between brain regions. The dynamic connectivity analyses suggest that interactions involving limbic regions vary over time in a manner that distinguishes AD from both MCI and healthy ageing. \par Previous studies have demonstrated that dynamic functional connectivity can reveal clinically relevant information by capturing fluctuations in connectivity occurring over shorter time scales \citep{rubido2024genetic,demirtacs2017whole}. Dynamic measures have been shown to improve the prediction of cognitive impairment in AD and to enhance discrimination between AD, MCI, and healthy individuals when compared with conventional static connectivity approaches \citep{du2018classification,quevenco2017memory}. In the present study, alterations observed in MCI were generally less pronounced and less consistent than those observed in AD. This likely reflects the biological heterogeneity of MCI, which includes individuals at different stages of disease progression and with different clinical trajectories. The ability of dynamic networks to capture fluctuations in functional interactions may therefore be particularly useful in identifying subtle changes that are not evident in time-averaged measures \citep{zhang2023static,gonzalez2018task}. \par 
Several observations emerged from the dynamic connectivity analyses. These included: (i) consistent inter-group differences involving connectivity between white matter regions and somatosensory cortices; (ii) the appearance of transient connections linking the Broca area and amygdala with the hippocampal formation; and (iii) the close correspondence between static and dynamic centrality measures, indicating that many white matter-related effects are local rather than network-wide in nature. \par 

The most consistent findings involved connectivity between white matter regions and somatosensory cortices (Figures~\ref{fig:static_link_ring_groups} and \ref{fig:sig_windows}). The interpretation of white-matter BOLD signals remains an active area of investigation. However, the view that white-matter fMRI signals primarily represent physiological noise does not fully reflect the current state of the field. Recent studies have demonstrated that white-matter BOLD fluctuations exhibit structured temporal dynamics, reproducible activation patterns, and functional interactions with distributed grey-matter networks \citep{schilling2023whole}. In addition, white-matter functional connectivity has been linked to both electrophysiological synchronisation and underlying structural connectivity, supporting its biological relevance \citep{huang2023intracranial}. Recent work across the Alzheimer's disease spectrum has further reported robust alterations in white-matter functional networks involving sensorimotor, frontal, and hippocampal systems \citep{itkyal2026investigating}. The white-matter effects observed in the present study were anatomically selective, reproduced across static and dynamic analyses, and remained evident across alternative parcellation schemes. Such spatially organised and biologically interpretable patterns would be unlikely if the observed signals were dominated by random noise. Our findings therefore support the growing body of evidence suggesting that white-matter functional signals contain meaningful information regarding large-scale brain organisation and its disruption in neurodegenerative disease. The involvement of somatosensory regions is also noteworthy. Although Alzheimer's disease has traditionally been viewed as relatively sparing primary sensory systems, recent work suggests that this interpretation may overlook substantial inter-individual variability in cognitive decline and sensory function \citep{wiesman2021somatosensory}. Cognitive domains commonly affected in AD, including attention and executive functioning, are known to influence somatosensory processing. Furthermore, recent studies have demonstrated interactions between somatosensory networks and prefrontal systems involved in cognitive control \citep{assem2024basis}. The observed alterations in white matter--somatosensory connectivity may therefore reflect broader disruptions of large-scale functional systems associated with disease progression rather than isolated abnormalities of sensory processing. The classification analyses support this interpretation. The highest discrimination between HC and AD was obtained using white-matter-derived features, yielding an AUC of 0.934. This suggests that connectivity patterns involving white matter regions contain meaningful disease-related information. Overall, our findings indicate that dynamic connectivity measures may provide complementary information to existing biomarker approaches based on structural imaging, molecular biomarkers, and cognitive assessment. \par 
Another notable finding was the emergence of transient connectivity involving the amygdala and hippocampal formation (Figs.~\ref{fig:nodestrength_groups}~and~\ref{fig:eig_groups}). The importance of amygdala-hippocampal interactions in Alzheimer's disease has been extensively documented \citep{song2023amygdala}. Disruption of this circuitry has been associated with impairments in episodic memory, emotional memory, and working memory, as well as with pathological processes including tau accumulation, amyloid deposition, synaptic dysfunction, and cholinergic deficits \citep{song2023amygdala}. The present results indicate that some of these alterations are more readily identified when temporal fluctuations in connectivity are considered, suggesting that dynamic analyses provide additional insight into disease-related changes within limbic networks. \par 
Our findings also complement recent reports highlighting the role of eigenvector centrality as a potential early indicator of Alzheimer's disease pathology and cognitive decline \citep{lorenzini2023eigenvector}. While node-strength analyses identified robust differences involving white matter regions, link-level analyses indicated that variability in centrality measures was largely associated with connectivity involving limbic structures, particularly the hippocampal formation and amygdala. These findings are consistent with previous studies identifying hippocampal-amygdala connectivity as a marker of early network dysfunction in Alzheimer's disease \citep{laakso1995volumes}. \par 
Several limitations should be acknowledged. First, the study was cross-sectional and therefore cannot determine how dynamic connectivity changes evolve over the course of disease progression. Second, although the ADNI cohort provides a well-characterised research dataset, it may not fully reflect the heterogeneity observed in routine clinical populations. Third, the classification analyses should be viewed as exploratory and were intended to assess the discriminative value of network-derived features rather than develop a clinically deployable clinical model. Finally, independent external validation will be required to establish the generalisability of these findings. The classification analyses were hypothesis-driven and intended to evaluate whether network alterations identified in the primary connectivity analyses possessed discriminative value at the individual level. \par
In conclusion, the present study demonstrates that dynamic functional connectivity provides information that complements conventional static connectivity analyses in Alzheimer's disease. The most consistent group differences involved interactions between white matter and somatosensory regions, whereas dynamic analyses additionally highlighted transient alterations involving amygdala-hippocampal circuitry and association networks. These findings suggest that disease-related network dysfunction involves not only changes in the strength of functional connections but also changes in their temporal organisation, supporting the value of dynamic connectivity approaches for investigating large-scale brain dysfunction across the Alzheimer's disease spectrum.

\section*{Author Contribution}
 N.R. conducted analysis on the fMRI data, prepared figures, and drafted the initial methods and results sections. V.B. performed the RF analysis, with M.F. contributing the same RF analysis on the dataset. V.V. supervised the project, conceptualised the study, and wrote the final manuscript.

 \section*{Funding Declaration}
N.R and M.F were supported by RSAT (Grant No. DSR1058-100), V.B. was funded by BRACE (Grant No. DSR1085-100).
\section*{Acknowledgements}

Data collection and sharing for this project was funded by the
Alzheimer's Disease Neuroimaging Initiative (ADNI)
(National Institutes of Health Grant U01 AG024904) and DOD ADNI
(Department of Defense award number W81XWH-12-2-0012).
ADNI is funded by the National Institute on Aging and the
National Institute of Biomedical Imaging and Bioengineering,
and through contributions from participating organisations.
The Canadian Institutes of Health Research provides funding
to support ADNI clinical sites in Canada.
Private sector contributions are facilitated by the
Foundation for the National Institutes of Health (FNIH).
The grantee organisation is the Northern California Institute
for Research and Education, and the study is coordinated by
the Alzheimer's Therapeutic Research Institute at the
University of Southern California.
ADNI data are disseminated by the Laboratory for Neuro Imaging
at the University of Southern California.

\section*{Data Availability Statement}

The data supporting the findings of this study are available
from the corresponding author upon reasonable request.
The original neuroimaging and clinical data were obtained
from the Alzheimer's Disease Neuroimaging Initiative (ADNI)
and are accessible to eligible researchers through the ADNI
data-sharing platform, subject to its data-use agreements.

\section*{Conflict of Interest}

The authors declare no conflicts of interest.

\clearpage
\appendix

\renewcommand{\thesection}{S\arabic{section}}
\renewcommand{\thetable}{S\arabic{table}}
\renewcommand{\thefigure}{S\arabic{figure}}
\setcounter{section}{0}
\setcounter{table}{0}
\setcounter{figure}{0}
\large{\section*{Supplementary Information}}
\section*{Data Analysis} \label{sec:dataanalysis}
\paragraph{Data Acquisition} We select high-definition T1-weighted (T1w) images and resting-state fMRI (rs-fMRI) data from the Alzheimer's Disease Neuroimaging Initiative 3 (ADNI-3) protocol \citep{weiner2017alzheimer}. T1w images are acquired using $3\,$Tesla scanners (GE, Siemens, or Philips), with 2 time acceleration, sagittaly (upper to lower), and using $1\,mm$ slices of $1\,mm^2$ pixels (Field-Of-View = $208\times240\times256mm$; Echo Time (TE) = min full echo $\sim3.0\,ms$; Time following Inversion Pulse (TI) = $900.0\,ms$; Repetition Time (TR) = $2300.0\,ms$, which approximately account to a total scan time of 6:20 minutes). rs-fMRI follow ADNI-3's basic EPI-BOLD protocol, where participants have their eyes open and are scanned for (nearly) $10$ minutes using the same two-time accelerated $3\,T$ scanners, following an even/odd interleaved axial-slicing (inferior to superior) of $3.4\,mm$ with $(3.4375\,mm)^2$ pixels (FOV = $220\times220\times163mm$; $P>>A$ phase encoding; TE = $30\,ms$; TR = $3000\,ms$). More details at \href{http://adni.loni.usc.edu/data-samples/data-types/mri}{ADNI's web-page}. 

\paragraph{Pre-processing Pipeline} Data pre-processing is done by FMRIB's Expert Analysis Tool \citep{woolrich2001temporal} (\texttt{FEAT}\footnote{URL: \href{https://fsl.fmrib.ox.ac.uk/fsl/fslwiki/FEAT/StepByStep1}{FSL's wiki on FEAT pre-processing} }), involving a high-pass filter (cut-off of at $100\,s$; $0.01\,Hz$), motion corrections (\texttt{MCFLIRT}) \citep{jenkinson2002improved}, inter-leaved slice-timing corrections (using Fourier-space signal's phase-shifting), non-brain removal (\texttt{BET}) \citep{smith2002fast}, spatial smoothing (using a Gaussian kernel of full-width-half-maximum $5\,mm$), and intensity normalisation. Such pre-processing is applied to all participants alike, holding filtered rs-fMRI in each participant's native space and with their original resolution (minimising data manipulation). Quality control is carried by a visual assessment of FEAT's outputs, where $3$ AD, $3$ MCI, and $2$ HC participants are discarded due to poor boundary-based registration (BBR) of rs-fMRI to T1w images. In addition, we excluded 31 subjects from the study whose motion-correction residuals were identified as outliers relative to the rest of the cohort. Subjects across all groups who exhibited excessive motion during fMRI preprocessing (6 in the AD group, 13 in the MCI group, and 12 in the HC group) were therefore removed from the analysis. Following quality control of the FEAT processing outputs, these subjects showed abnormally high voxel-wise synchrony across the entire brain, indicating that motion correction was insufficient to adequately compensate for head movement during fMRI acquisition. A list of the subjects can be found in the OFS repository folder of this project.

\paragraph{Brain Atlas Registration.} Registration of the atlas (from MNI space) to each participant's rs-fMRI native space is done by FMRIB's Linear Image Registration Tool \citep{jenkinson2002improved, jenkinson2001global} (FLIRT). FLIRT transforms the $2\,mm$ resolution atlas in MNI space to the resolution of the filtered rs-fMRI ($3.4375 \times 3.4375 \times 3.4\,mm$) in native space orientation. The necessary matrix transformation for this registration is obtained from FEAT's rs-fMRI registration step. This step involves a BBR of the rs-fMRI to the participant's T1w high-resolution brain-extracted image, which we obtain by applying the \texttt{fsl\_anat}\footnote{URL: \href{https://fsl.fmrib.ox.ac.uk/fsl/fslwiki/fsl_anat}{FSL's wiki on fsl\_anat} } (default) pipeline, and a registration to the standard MNI152 brain. For the JBA regions' list see Table~\ref{tab:JA_regions}.

\begin{table*}
\begin{center}
\noindent\begin{minipage}{0.48\linewidth}
\begin{tabular}{@{}|l r|@{}}
\hline
Indices & Region  \\
\hline
1/2 & Anterior intra-parietal sulcus hIP1 L/R \\
3/4 & Anterior intra-parietal sulcus hIP2 L/R\\
5/6&Anterior intra-parietal sulcus hIP3 L/R\\
7/8& Amygdala centromedial group L/R\\
9/10 & Amygdala laterobasal group L/R\\
11/12& Amygdala superficial group L/R\\
13/14& Broca's area BA44 L/R\\
15/16&Broca's area BA45 L/R\\
17/18&Hippocampus cornu ammonis L/R\\
19/20&Hippocampus entorhinal cortex L/R\\
21/22& Hippocampus dentate gyrus L/R\\
23/24& Hippocampal-amygdaloid transition area L/R\\
25/26&Hippocampus subiculum L/R\\
27/28&Inferior parietal lobule PF L/R\\
29/30& Inferior parietal lobule PFcm L/R\\
31/32&Inferior parietal lobule PFm L/R\\
33/34&Inferior parietal lobule PFop L/R\\
35/36& Inferior parietal lobule PFt L/R\\
37/38&Inferior parietal lobule Pga L/R\\
39/40&Inferior parietal lobule PGp L/R\\
41/42&Primary auditory cortex TE1.0 L/R\\
43/44&Primary auditory cortex TE1.1 L/R\\
45/46&Primary auditory cortex TE1.2 L/R\\
47/48 &  Primary motor cortex BA4a L/R\\
49/50 &  Primary motor cortex BA4p L/R\\
51/52& Primary somatosensory cortex BA1 L/R\\
53/54&  Primary somatosensory cortex BA2 L/R\\
55/56&  Primary somatosensory cortex BA3a L/R\\
57/58&  Primary somatosensory cortex BA3b L/R\\
59/60 & Sec som cortex/Parietal operculum OP1 L/R\\
61/62&  Secondary SC/OP2 L/R\\
\hline
\end{tabular}
\end{minipage}
\hfill
\noindent\begin{minipage}{0.49\linewidth}
\centering
\begin{tabular}{|l r|}
\hline
Indices & Region  \\
\hline
63/64 & Secondary SC/OP3 L/R\\
65/66&  Secondary SC/OP4 L/R\\
67/68 &  Superior parietal lobule 5Ci L/R\\
69/70 &  Superior parietal lobule 5L L/R\\
71/72 &  Superior parietal lobule 5M L/R\\
73/74 &  Superior parietal lobule 7A L/R\\
75/76 &  Superior parietal lobule 7M L/R\\
77/78&  Superior parietal lobule 7PC L/R\\
79/80 &  Superior parietal lobule 7P L/R\\
81/82 & Visual cortex V1 BA17 L/R\\
83/84 &  Visual cortex V2 BA18 L/R\\
85/86 & Visual cortex V3V L/R\\
87/88 &  Visual cortex V4 L/R\\
89/90& Visual cortex V5 L/R\\
91/92&  Premotor cortex BA6 L/R\\
\hline
93/94 & Acoustic radiation R/L\\
95 &  Callosal body\\
96/97 &  Cingulum R/L\\
98/99 &  Corticospinal tract R/L\\
100 &  Fornix\\
101/102 &  Inferior occipito-frontal fascicle R\/L\\
103/104 &  Lateral geniculate body R/L\\
105 &  Mamillary body\\
106/107 & Medial geniculate body R/L\\
108/109 & Optic radiation R/L\\
110/111 &  Superior longitudinal fascicle R/L\\
112/113 &  Superior occipito-frontal fascicle R/L\\
114/115 &  Uncinate fascicle R/L\\
\hline
116/117 &  Insula Id1 L/R\\
118/119 &  Insula Ig1 L/R\\
120/121 &  Insula Ig2 L/R\\
\hline
\end{tabular}
\end{minipage}
\end{center}
\caption{Juelich Brain Atlas regions. The mid lines indicate border between grey and white matter regions. In the main text, we also use SSC and PSC to denote the secondary and primary somatosensory cortices, respectively.}
\label{tab:JA_regions}
\end{table*}

\begin{figure*}
    \centering
    \includegraphics[width=0.31\linewidth]{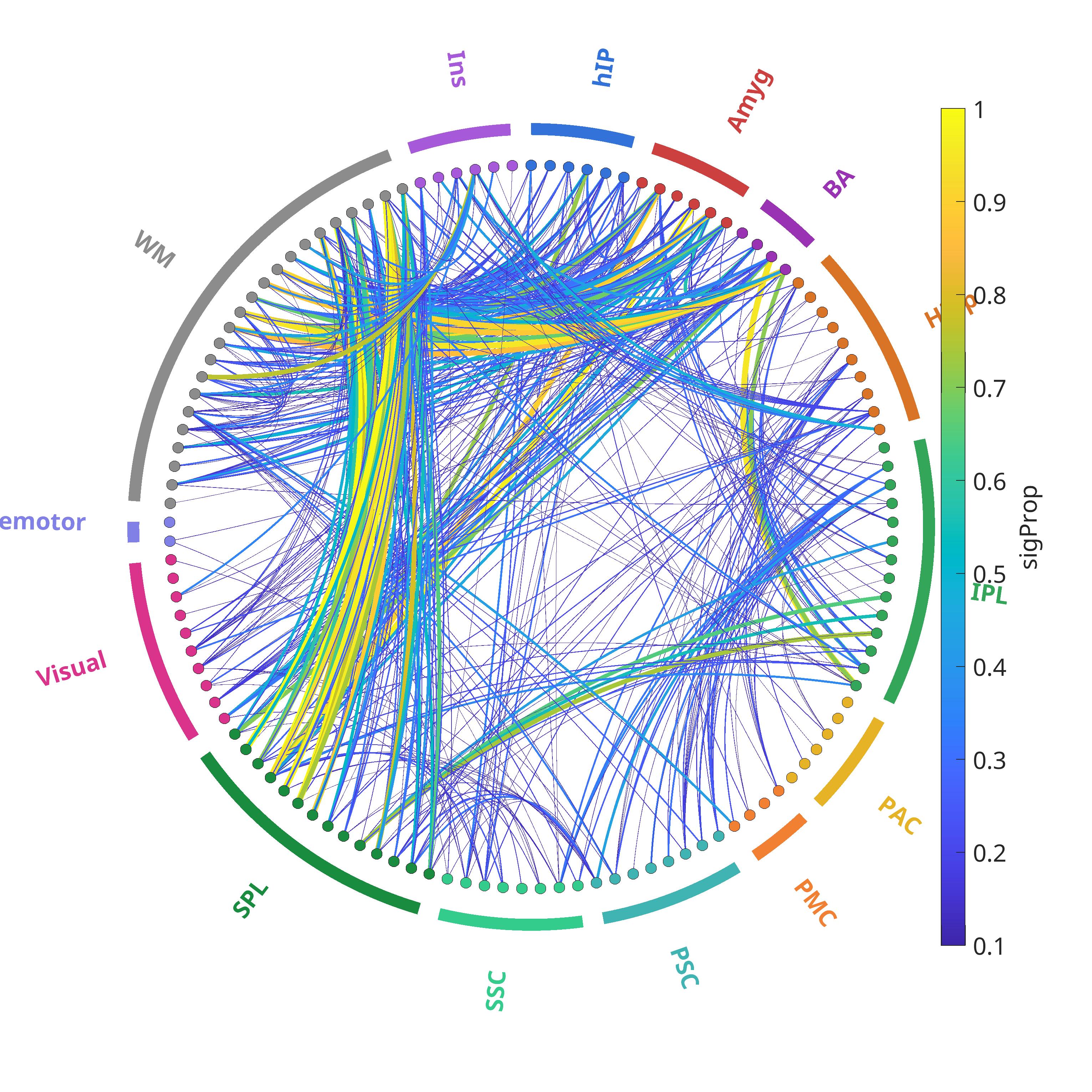}
    \includegraphics[width=0.31\linewidth]{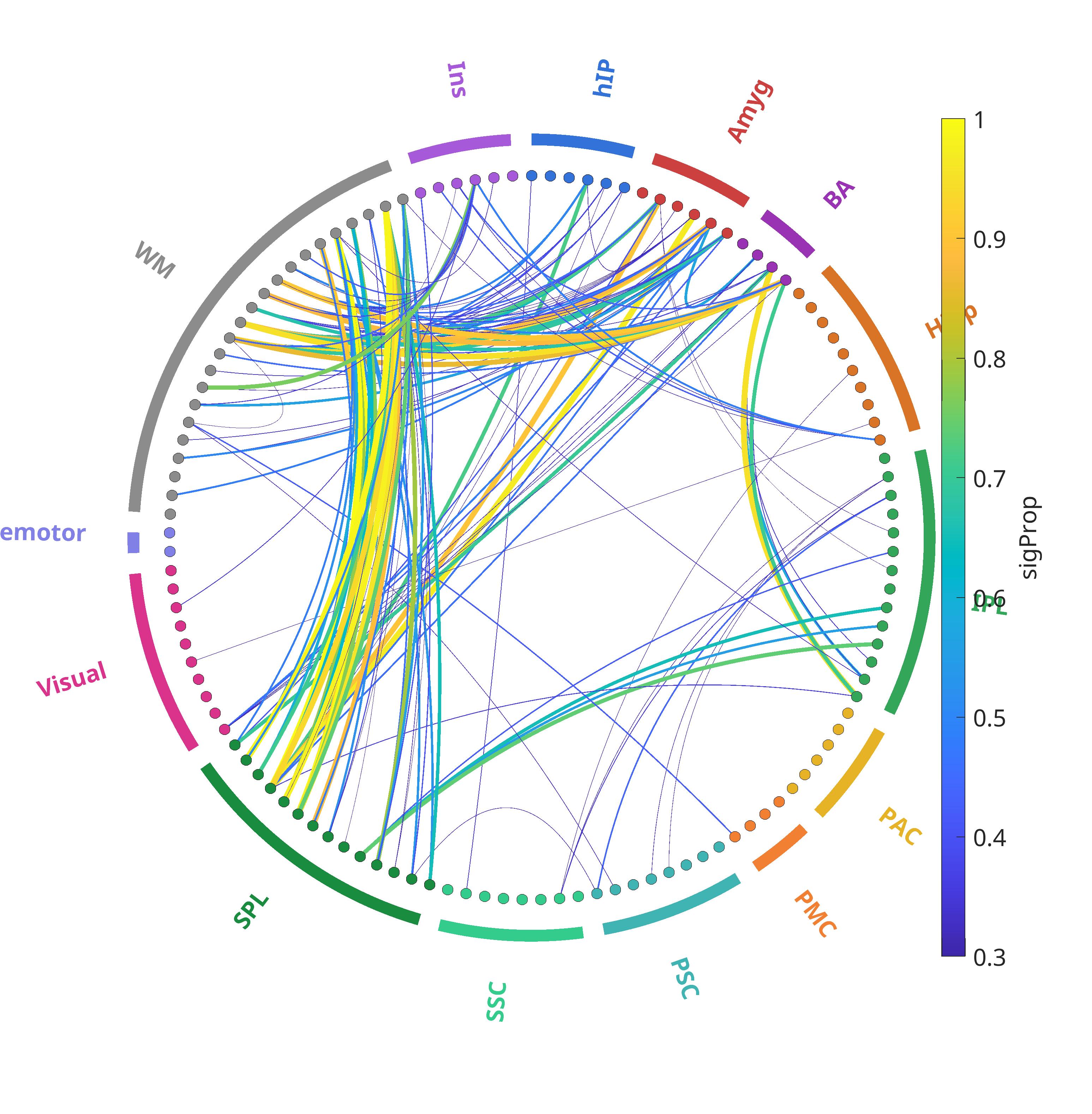}
    \includegraphics[width=0.31\linewidth]{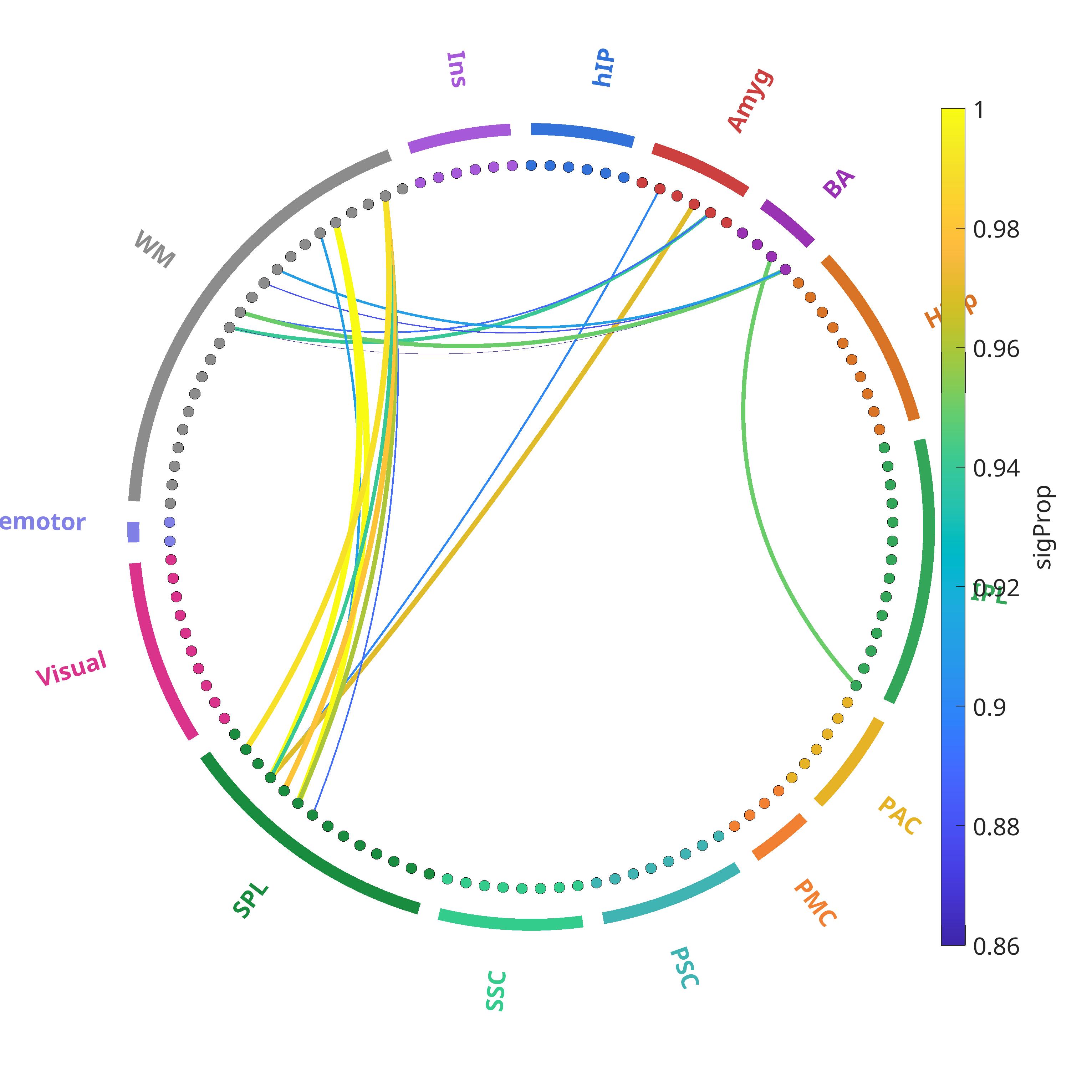}
    \caption{\textbf{Static functional network differences at the link level across the three study groups.} Links (i.e., pairwise correlations) between nodes of the J{\"u}lich Brain Atlas are shown when they pass the Kruskal–Wallis (K-W) tests for group differences (HC, MCI, AD). For each node pair $(i,j)$, statistical significance was evaluated using 100 independent resampling realisations, each consisting of 150 participants (50 per group). Colour coding represents the portion of each link passed the K-W test ($p<0.01$). Significant portion of links survived $10\%$ (Left), $33\%$ (Middle) and $80\%$ (Right) bootstraps. This procedure ensures that reported differences reflect robust group effects across balanced cohort samples. A full list of regions, their labels, indices and abbreviations can be found in Supplementary Table~\ref{tab:JA_regions}.}
    \label{figsi:ring_KW_sigProp_thresholds}
\end{figure*}

\subsection*{Harvard-Oxford Brain Atlas}
\begin{figure*}
    \centering
    \includegraphics[width=0.31\linewidth]{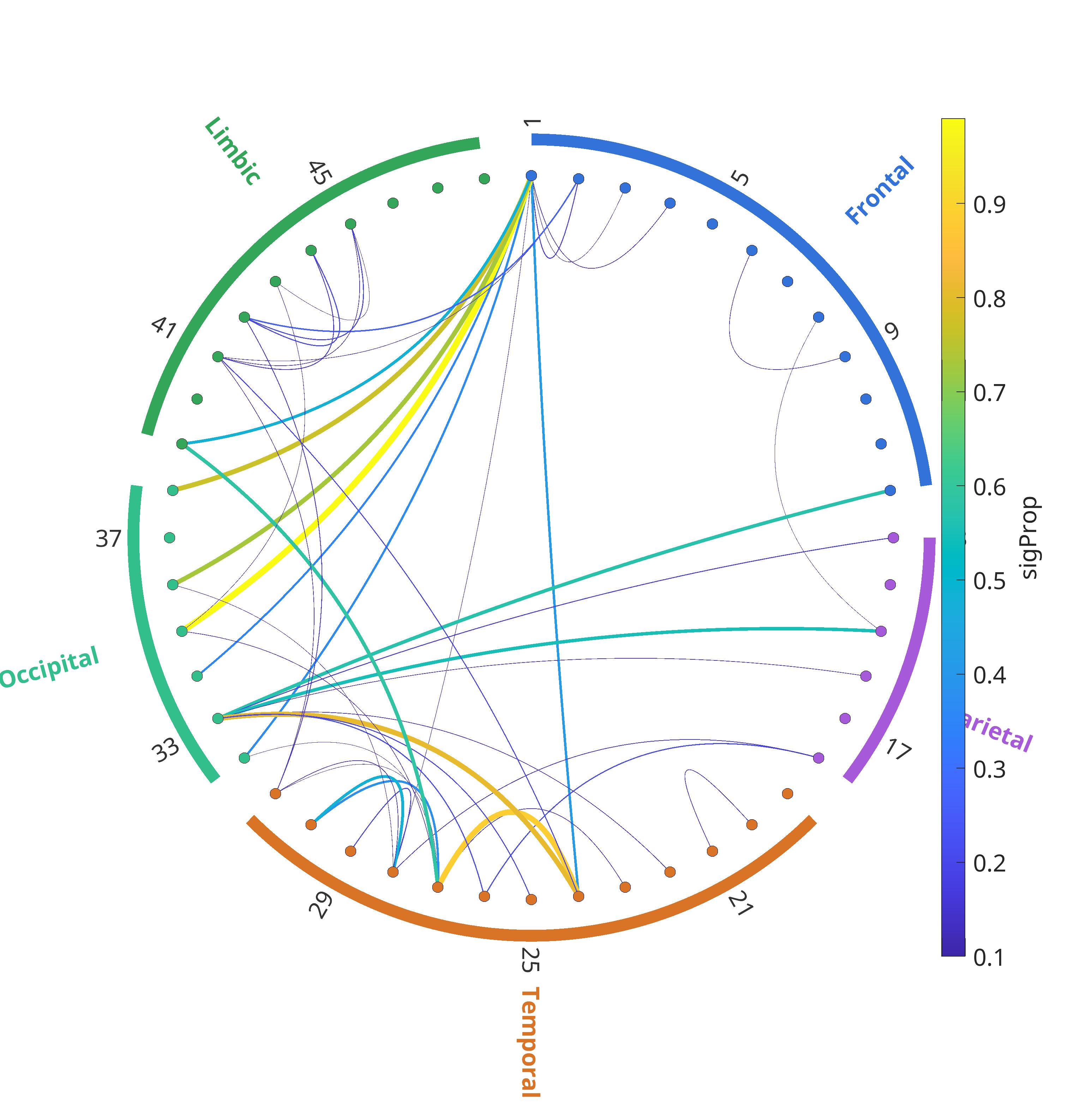}
    \includegraphics[width=0.31\linewidth]{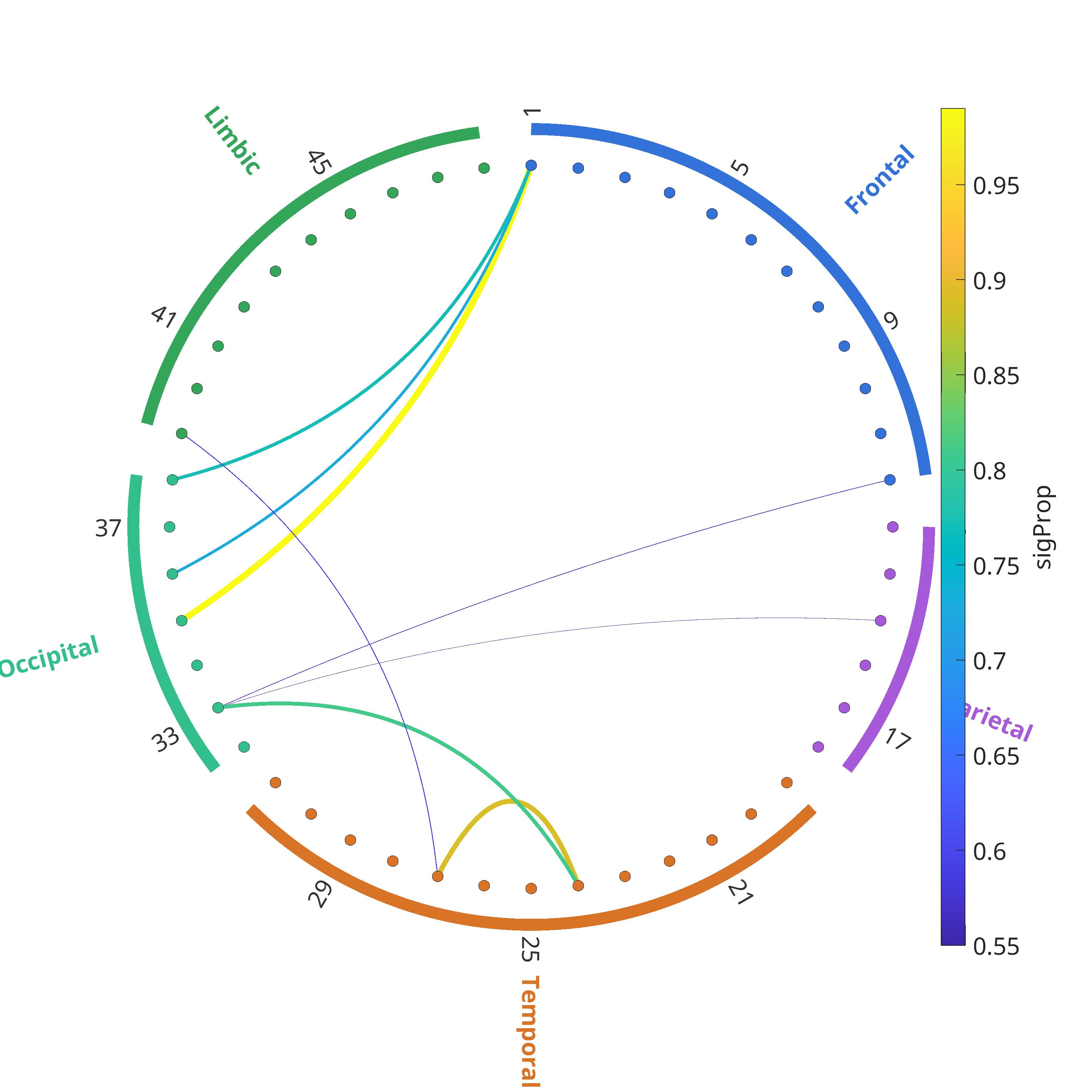}
    \includegraphics[width=0.31\linewidth]{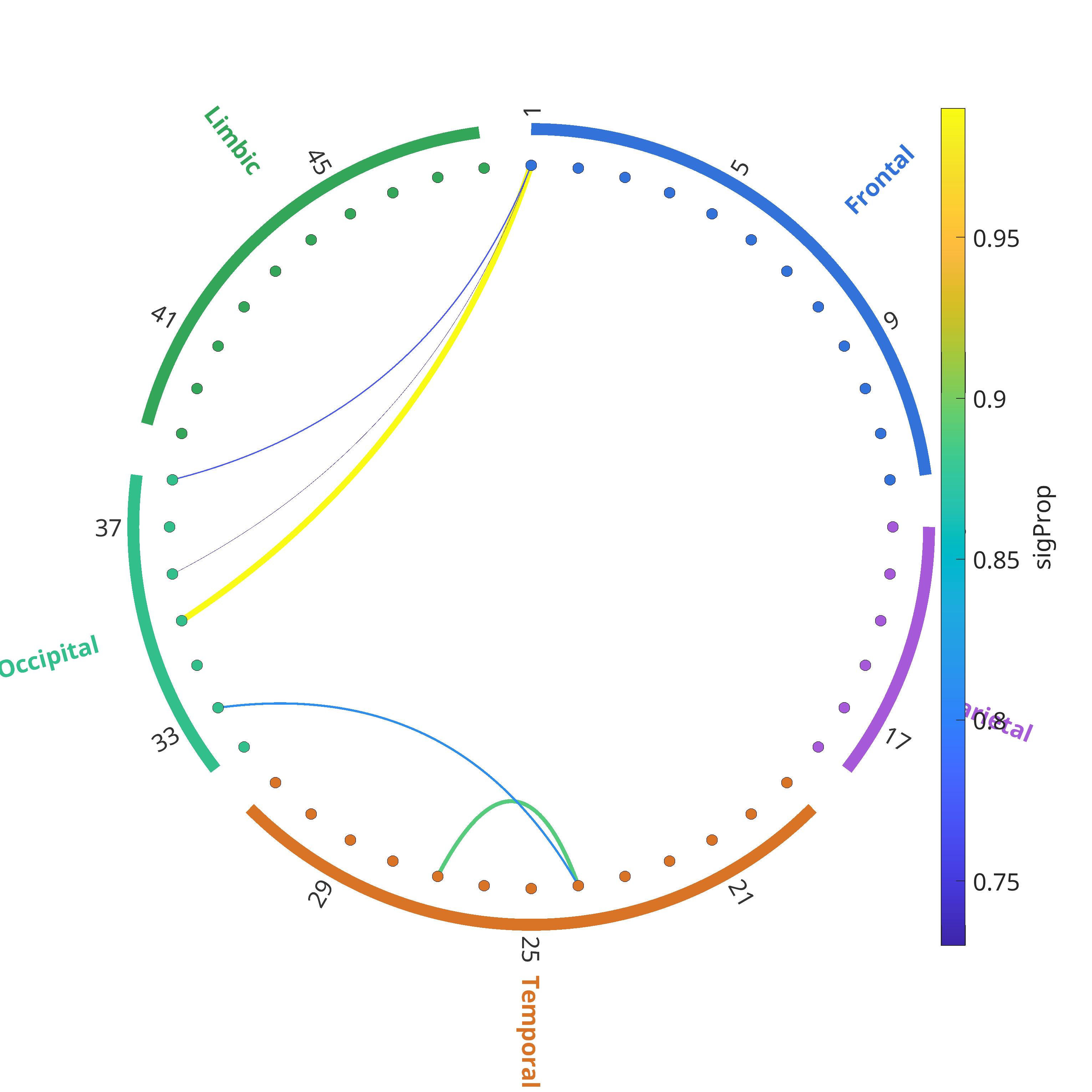}
    \caption{\textbf{Static functional network differences at the link level across the three study groups.} Links (i.e., pairwise correlations) between nodes of the Harvard-Oxford Brain Atlas are shown when they pass the Kruskal–Wallis (K-W) tests for group differences (HC, MCI, AD). For each node pair $(i,j)$, statistical significance was evaluated using 100 independent resampling realisations, each consisting of 150 participants (50 per group). Colour coding represents the portion of each link passed the K-W test ($p<0.01$). Significant portion of links survived $10\%$ (Left), $50\%$ (Middle) and $80\%$ (Right) bootstraps. This procedure ensures that reported differences reflect robust group effects across balanced cohort samples. Colour-coded is nodal lobar affiliation. The analysis includes $48$ cortical nodes of the H-OBA.}
    \label{figsi:ring_KW_sigProp_HOBA}
\end{figure*}

An independent cortical parcellation was done based on the H-OBA 48 regions. Results of the group-wise analysis using the K-W test (similar to Fig.~\ref{figsi:ring_KW_sigProp_thresholds} for the JBA) are shown in Fig.~\ref{figsi:ring_KW_sigProp_HOBA}.

\section*{Random Forest Classification}
\begin{figure*}[h!]
    \centering
\includegraphics[width=0.24\linewidth]{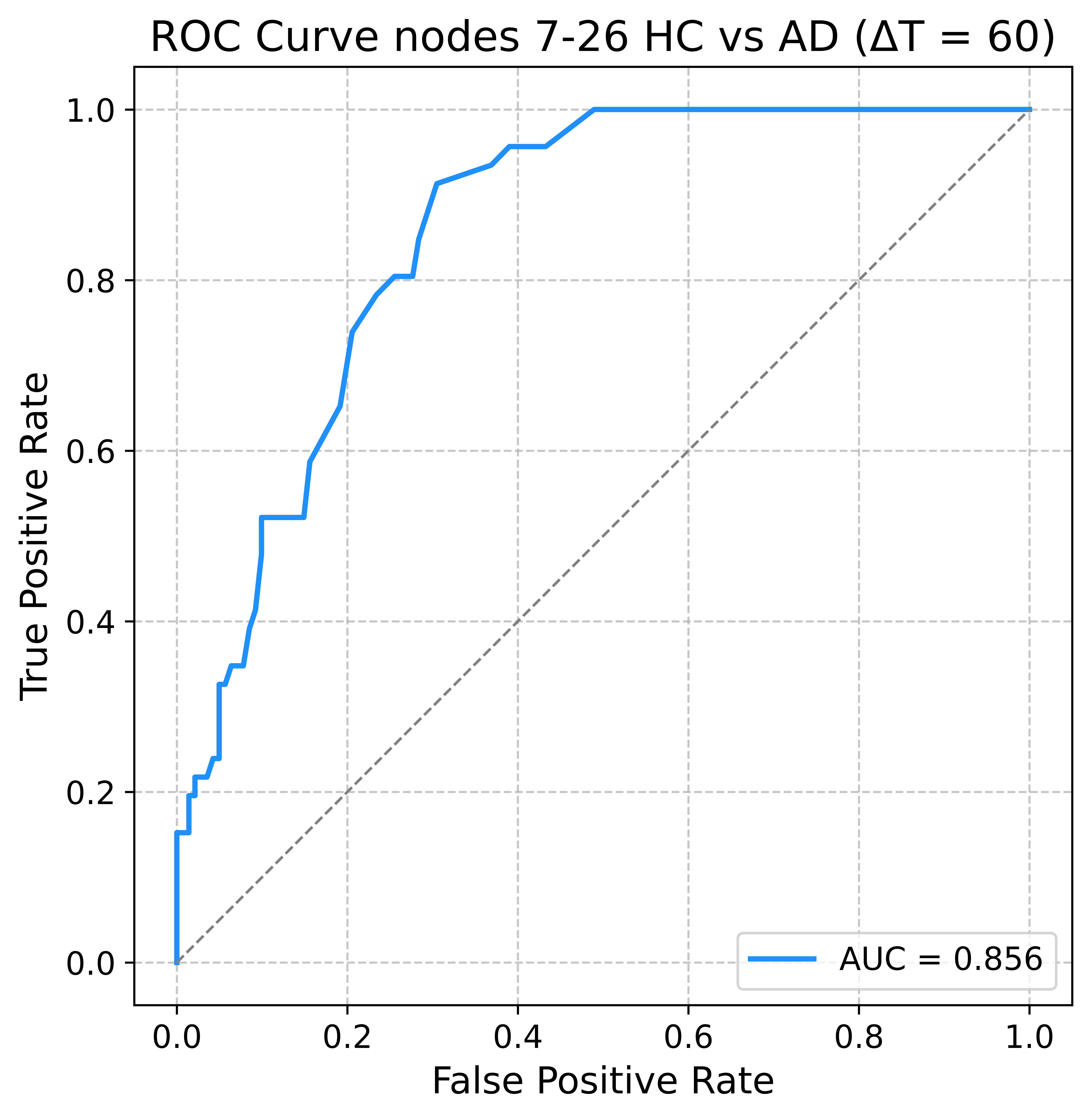}
\includegraphics[width=0.24\linewidth]{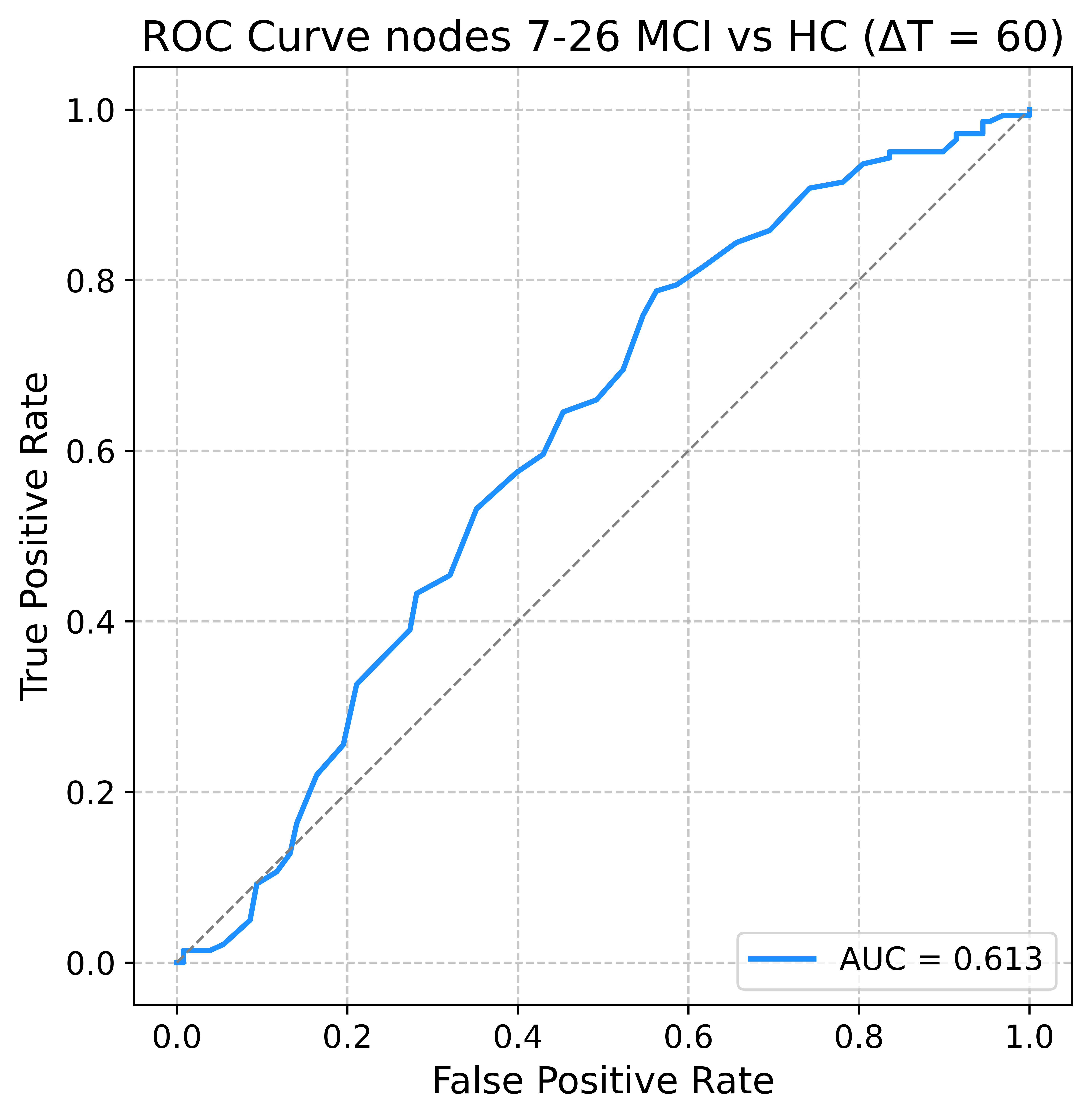}
\includegraphics[width=0.24\linewidth]{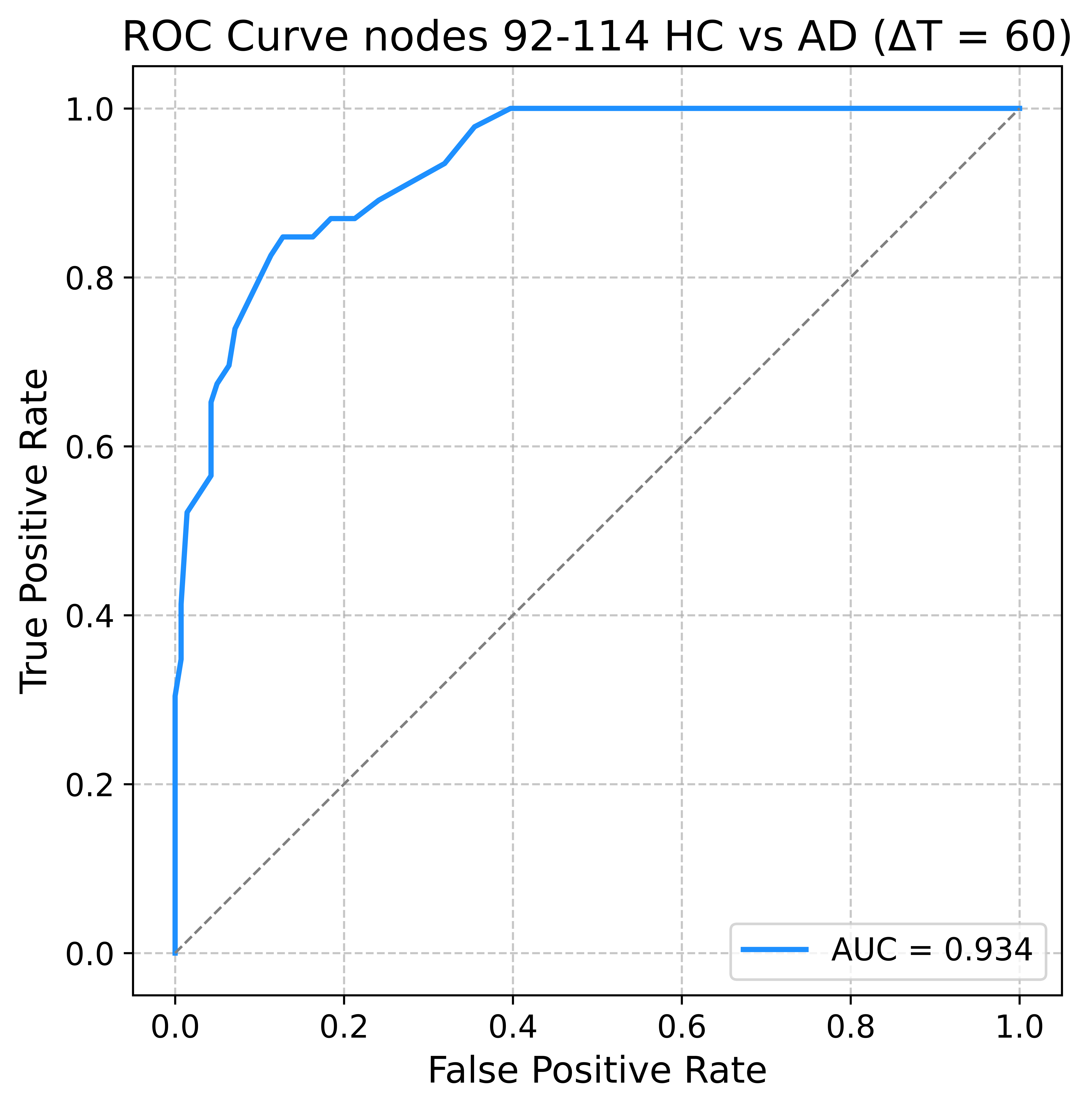}
\includegraphics[width=0.25\linewidth]{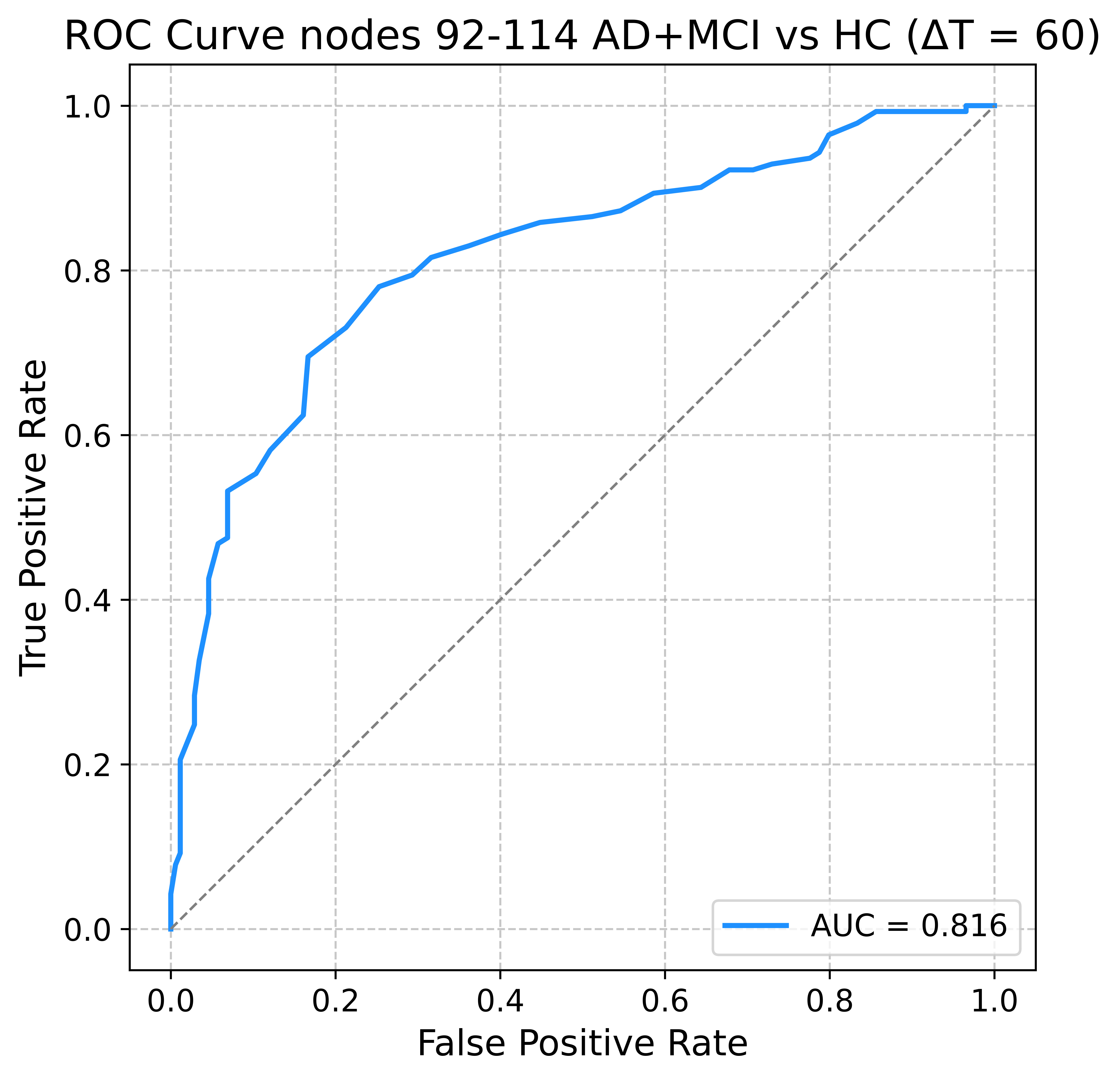}
\caption{\textbf{Receiver operating characteristic (ROC) curves for Random Forest classification using dynamic BOLD time series at $\Delta t = 180$ s}. From left to right: Pairwise classification using nodes 7--26 showed good HC versus AD discrimination (AUC = 0.856), moderate MCI versus AD discrimination (AUC = 0.760), and weaker HC versus MCI discrimination (AUC = 0.630). 
Far right: Classification using nodes 93--113 distinguished AD+MCI from HC with AUC = 0.816.
Please note: $\Delta T$ = 60 time points = 180 s.}
\label{figsi:ROC}
\end{figure*}
To further evaluate the clinical utility of the identified network-derived regions, receiver operating characteristic (ROC) analyses were performed for the best-performing dynamic connectivity models ($\Delta t = 180$ s, i.e., $\Delta T = 60$ in the figure). For the limbic node set (nodes 7--26), classification performance was highest for HC versus AD (AUC = 0.856), indicating good discrimination. Classification performance was lower for MCI versus AD (AUC = 0.760) and HC versus MCI (AUC = 0.630), consistent with the intermediate and heterogeneous nature of MCI. For the white-matter node set (nodes 93--113), the highest classification performance was observed for HC versus AD (AUC = 0.934). This finding indicates that dynamic functional interactions involving white-matter regions contribute substantially to the separation of Alzheimer's disease participants from healthy controls.

\subsection*{Dynamic Networks Between Groups: Deviation from Healthy Controls} \label{sec:direction-dFN-median} 
\begin{figure}
    \centering
    \includegraphics[width=0.45\linewidth]{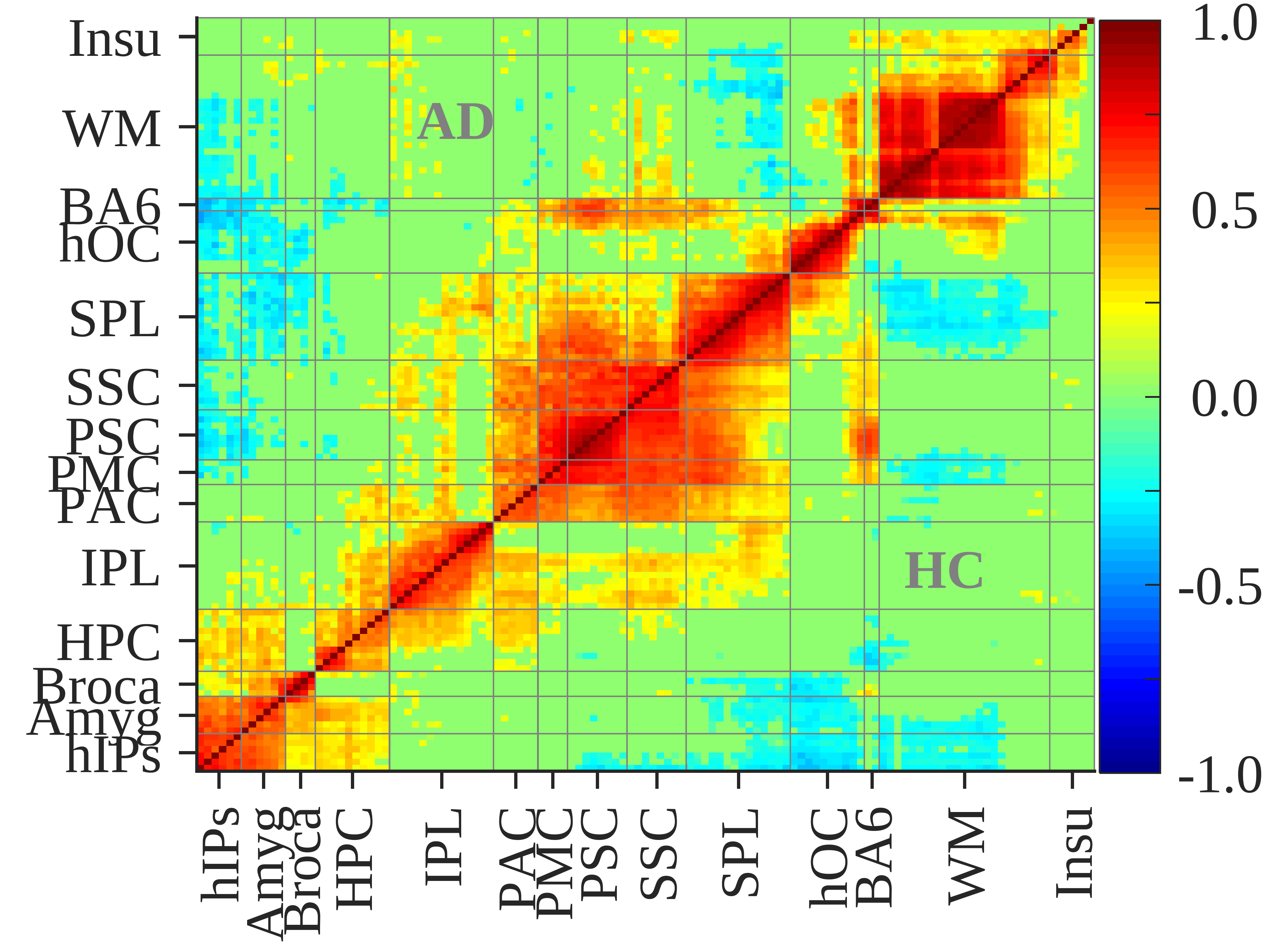}
    \includegraphics[width=0.45\linewidth]{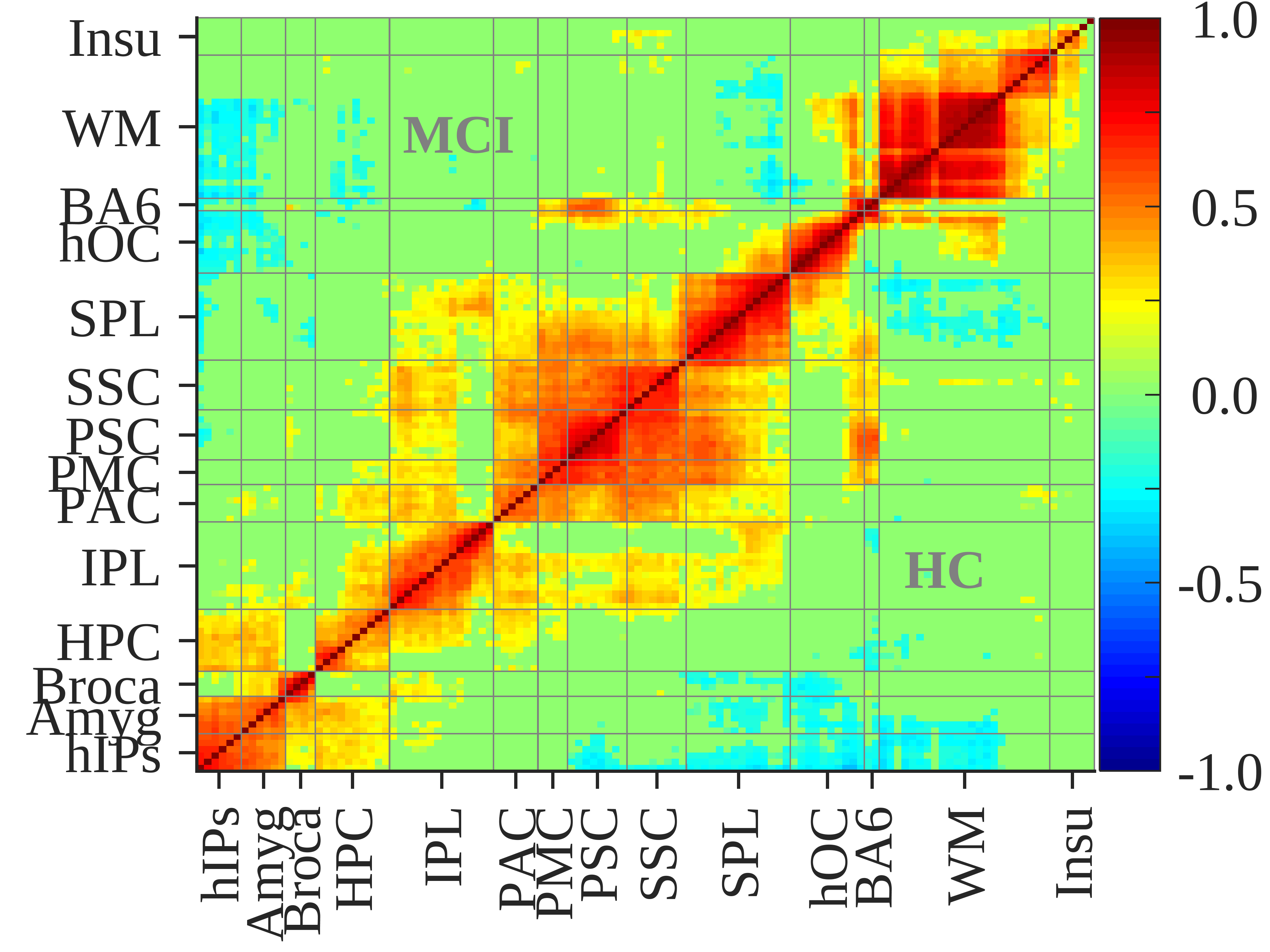}
    \caption{\textbf{Cohorts’ median dynamic correlations.} Upper diagonal entries show the median dynamic correlation values
(color-coded) for the AD (left) and MCI (right) cohort, and lower diagonal entries show the corresponding values for the
HC cohort. Cohort labels (AD, MCI, and HC) are displayed within the matrices to indicate the corresponding upper and
lower triangular sections. Thin black lines delineate groups of anatomically related regions. Black thin, lines separate
regions within similar anatomical landmarks. Abbreviations: Alzheimer’s disease (AD), mild cognitive impairment
(MCI) and healthy controls (HC). A full list of regions, including their labels, indices, and abbreviations, is provided in
Supplementary Table S1.}
    \label{fig:admci_hc_matrices}
\end{figure}

Figure~\ref{fig:admci_hc_matrices} presents group-level median dynamic functional networks. In contrast to the static networks, connectivity differences varied across sliding windows, reflecting temporal fluctuations in functional interactions. The most consistent alterations involved connections between white-matter and parietal regions. In particular, connectivity between the superior parietal lobule (SPL) and white matter was reduced in AD relative to HC, whereas interactions between the inferior parietal lobule (IPL) and white matter emerged preferentially in AD. Additional transient differences were observed in limbic networks, including connections between the amygdala and primary sensory cortex (PSC). These findings indicate that dynamic analyses reveal temporally varying connectivity alterations that are less apparent in time-averaged static networks.

\end{document}